\documentclass[12pt,letterpaper]{article}
\pdfoutput=1

\usepackage{amsmath,amssymb,calc, amsthm,bbm, epsfig,psfrag, mathtools}
\usepackage{graphicx, enumerate}
\usepackage[numbers,sort&compress]{natbib}
\usepackage[dvipsnames]{xcolor}
\usepackage{tensor}
\usepackage{empheq}
\usepackage{fancybox}

\usepackage[utf8]{inputenc} 

\newcommand{\Comment}[1]{{}}
\definecolor{darkblue}{rgb}{0.15,0.35,0.55}
\definecolor{reddish}{rgb}{0.65, 0.2, 0.2}

\usepackage[linktocpage=true]{hyperref}
\hypersetup{
colorlinks=true,
citecolor=darkblue,
linkcolor=reddish,
urlcolor=darkblue,
pdfauthor={},
pdftitle={},
pdfsubject={}
}

\usepackage{mathrsfs}

\PassOptionsToPackage{linecolor=blue,backgroundcolor=blue!25,bordercolor=blue,textsize=scriptsize}{todonotes}
\usepackage{todonotes}

\makeatletter
\renewcommand\section{\@startsection {section}{1}{\z@}%
                                   {-3.5ex \@plus -1ex \@minus -.2ex}
                                   {2.3ex \@plus.2ex}%
                                   {\normalfont\large\bfseries}}
\renewcommand\subsection{\@startsection{subsection}{2}{\z@}%
                                     {-3.25ex\@plus -1ex \@minus -.2ex}%
                                     {1.5ex \@plus .2ex}%
                                     {\normalfont\bfseries}}
\makeatother

\let\non\nonumber

\def\AdS{{\rm AdS}}

\def\bea#1\eea{\begin{align}#1\end{align}}

\def\bes #1\ees{\begin{split}#1\end{split}}

\newcommand{\be}{\begin{equation}}
\newcommand{\ee}{\end{equation}}

\newcommand{\bma}{\begin{pmatrix}}
\newcommand{\ema}{\end{pmatrix}}

\newcommand{\VT}{{\widetilde{V}_{4}}}

\newcommand{\hn}{\hat{n}}

\newcommand{\Z}{{\mathbb Z}}
\newcommand{\R}{{\mathbb R}}

\let\a=\alpha

\let\S=\Sigma

\def\e{\epsilon}

\def\m{\mu}
\def\n{\nu}

\def\M{{\mathcal M}}

\newcommand{\p}{\partial}

\newcommand{\C}[1]{$(\ref{#1})$}

\def\IZ{\relax\ifmmode\mathchoice
{\hbox{\cmss Z\kern-.4em Z}}{\hbox{\cmss Z\kern-.4em Z}}
{\lower.9pt\hbox{\cmsss Z\kern-.4em Z}} {\lower1.2pt\hbox{\cmsss
Z\kern-.4em Z}}\else{\cmss Z\kern-.4em Z}\fi}
\def\IR{\relax{\rm I\kern-.18em R}}

\def\one{{\hbox{ 1\kern-.8mm l}}}

\newlength{\bredde}
\def\slash#1{\settowidth{\bredde}{$#1$}\ifmmode\,\raisebox{.15ex}{/}
\hspace*{-\bredde} #1\else$\,\raisebox{.15ex}{/}\hspace*{-\bredde}
#1$\fi}

\newsavebox{\zzzbar}
\sbox{\zzzbar}
  {\setlength{\unitlength}{0.9em}
  \begin{picture}(0.6,0.7)
  \thinlines
  \put(0,0){\line(1,0){0.6}}
  \put(0,0.75){\line(1,0){0.575}}
  \multiput(0,0)(0.0125,0.025){30}{\rule{0.3pt}{0.3pt}}
  \multiput(0.2,0)(0.0125,0.025){30}{\rule{0.3pt}{0.3pt}}
  \put(0,0.75){\line(0,-1){0.15}}
  \put(0.015,0.75){\line(0,-1){0.1}}
  \put(0.03,0.75){\line(0,-1){0.075}}
  \put(0.045,0.75){\line(0,-1){0.05}}
  \put(0.05,0.75){\line(0,-1){0.025}}
  \put(0.6,0){\line(0,1){0.15}}
  \put(0.585,0){\line(0,1){0.1}}
  \put(0.57,0){\line(0,1){0.075}}
  \put(0.555,0){\line(0,1){0.05}}
  \put(0.55,0){\line(0,1){0.025}}
  \end{picture}}

\newfont{\goth}{ygoth.tfm scaled 1200}                   

 \numberwithin{equation}{section}

\def\1{{(1)}}
\def\2{{(2)}}
\def\3{{(3)}}

\newcommand{\overbar}[1]{\mkern 1.5mu\overline{\mkern-1.5mu#1\mkern-1.5mu}\mkern 1.5mu}

\def\TT{{T\overbar{T}}}

\newcommand{\ul}{\underline}

\begin{document}
\begin{titlepage}

\begin{center}

\today
\hfill         \phantom{xxx}  EFI-23-09

\vskip 2 cm {\Large \bf Holography with Null Boundaries} 
\vskip 1.25 cm {\bf Christian Ferko$^{1, 2}$ and Savdeep Sethi$^{3}$}\non\\

\vskip 0.2 cm
 {\it $^1$ Department of Physics, Northeastern University, Boston, MA 02115, USA}

 \vskip 0.2 cm
 {\it $^2$ The NSF Institute for Artificial Intelligence
and Fundamental Interactions}

 \vskip 0.2 cm
{\it $^3$   Leinweber Institute for Theoretical Physics \& Enrico Fermi Institute \& Kadanoff Center for Theoretical Physics\\ University of Chicago, Chicago, IL 60637, USA}

\end{center}
\vskip 1.5 cm

\begin{abstract}
\baselineskip=18pt

One of the key issues in holography is going beyond $\AdS$ and defining quantum gravity in spacetimes with a null boundary. Recent examples of this type involve linear dilaton asymptotics and are related to the $\TT$ deformation. 
We present a holographic correspondence derived from string theory, which is an example of a kind of celestial holography. The holographic definition is a spacetime non-commutative open string theory supported on D1-D5 branes together with fundamental strings. The gravity solutions interpolate between $\AdS_3$ metrics and six-dimensional metrics. Radiation can escape to null infinity, which makes both the encoding of quantum information in the boundary and the dynamics of black holes quite different from $\AdS$ spacetimes.

\end{abstract}

\end{titlepage}

\tableofcontents

\section{Introduction} \label{intro}
\subsubsection*{\ul{\it Motivation}}

The aim of this work is to study quantum gravity in a family of spacetimes obtained from string theory which admit a holographic definition. The best known example of holography is the $\AdS$/CFT correspondence where conformal field theory provides a definition of quantum gravity in spacetimes which are asymptotically $\AdS$. In this work we are interested in spacetimes which are not asymptotically $\AdS$. Rather the spacetimes of interest to us allow radiation to escape to null infinity. By contrast, radiation in $\AdS$ spacetimes can reach the timelike boundary in finite time and reflect. This is why large $\AdS$ black holes can exist as stable states in equilibrium with a radiation bath of their own Hawking quanta. This is not the case for the spacetimes we will describe and the dynamics of black holes is correspondingly quite different.

The holographic dual for these spacetimes with null boundary is interesting: it is not a local quantum field theory. Rather it is a non-commutative field theory with spacetime non-commutativity. Such theories were studied in, for example~\cite{Gopakumar:2000na,Seiberg:2000ms,Ganor:2000my,Gomis:2000zz, Robbins:2003ry}, with the realization that in certain limits they are actually non-commutative open string theories (NCOS)~\cite{Gopakumar:2000na,Seiberg:2000ms}. These are open string theories which do not give rise to a massless graviton. At low energies, they can loosely be viewed as field theories deformed by a very specific infinite collection of irrelevant operators that can be organized in the form of a Moyal or star product. The product takes the following form between two functions $f(x)$ and $g(x)$ with non-commutativity parameter $\theta^{ij}$,
\begin{align}
     f(x) \star g(x) = e^{ \frac{i}{2} \theta^{ij} {\p_i^{(1)}} {\p_j^{(2)}}} f(x_1) g(x_2)\vert_{x_1=x_2=x} \, .
\end{align}
In string theory, $\theta^{ij}$ is determined by a choice of NS $B_2$-field on the world-volume of a collection of D-branes.\footnote{One can also study their various U-dual variants~\cite{Gopakumar:2000ep,Cai:2000yk,Russo:2000zb}.} For a review of non-commutative field theory see, for example,~\cite{Douglas:2001ba}. There was subsequently a great deal of work constructing gravity solutions for various Dp-branes with spacetime non-commutativity; see, for example,~\cite{Maldacena:1999mh,Harmark:1999xt,Harmark:1999rb,Alishahiha:1999ci,Harmark:2000wv,Harmark:2000ff,Berman:2000jw,Russo:2000mg,Dasgupta:2003us}.

Most constructions of this type break Lorentz invariance by the background $B_2$-field. However, the theory on D1-branes is special because the $B_2$-field preserves Lorentz invariance. This NCOS theory, studied in~\cite{Gukov:1997tt,Gopakumar:2000ep,Klebanov:2000pp,Herzog:2000sr,Sahakian:2000ay}, can be regarded as a limit of a bound state of D1-branes and fundamental strings. A connection between NCOS and nonrelativistic limits was explored in~\cite{Guijosa:2023qym}.

Our interest is actually in a variant that was not studied in the past. We start with the D1-D5 system and turn on a spacetime $B_2$-field along the direction of the D1-strings. Equivalently, this is a bound state of D1-strings, D5-branes and F1-strings (fundamental strings) studied in a specific decoupling limit.

The advantage of the D1-D5-F1 case is that the interior is described by the holographic CFT defining the D1-D5 fully decoupled spacetime. So the metrics we find interpolate between an $\AdS_3$ spacetime in the interior and a $D=6$ spacetime near infinity. The reason for the change in dimension is that the NCOS decoupling limit retains the dynamics on the D5-branes, which dominate the asymptotics. We will summarize our findings momentarily.  

\subsubsection*{\ul{\it Relation to ALD spacetimes and celestial holography}}

One of our original motivations for considering spacetime non-commutativity was a talk by G.~Lechner~\cite{Lechner2019Deformations}, based on~\cite{Lechner:2011tq} and further explored in~\cite{Shyam:2022iwd},  which described a  mysterious connection between spacetime non-commutativity and the $\TT$ deformation of two-dimensional quantum field theory~\cite{Zamolodchikov:2004ce,Cavaglia:2016oda}; for reviews of the $\TT$ deformation, see \cite{Jiang:2019hxb,He:2025ppz}. Understanding this proposed connection led us to revisit NCOS theories in $1+1$-dimensions. 

In  string theory, 
there has been recent work on three-dimensional asymptotically linear dilaton spacetimes (ALD); see, for example~\cite{Forste:1994wp,Israel:2003ry,Giveon:2017nie,Giveon:2017myj,Asrat:2017tzd,Araujo:2018rho,Apolo:2019zai,Chang:2023kkq,Fichet:2023xbu}.  These spacetimes, obtained from boundstates of F1-strings and NS5-branes, interpolate between $\AdS_3$ metrics and linear dilaton metrics near infinity. Like the case studied here, radiation in these spacetimes can also escape to null infinity. Unlike the cases studied here, the boundary always remains two-dimensional. 

The holographic definition for the NS5-F1 string case involves a sector of little string theory. Intriguingly, the holographic theory shares some properties with a symmetric product CFT deformed by a single-trace $\TT$-deformation. However, this is not precisely true aside from the special case of the tensionless theory~\cite{Dei:2024sct}. Only in that case is the undeformed CFT actually a symmetric product, which can admit a single-trace $\TT$ deformation. Yet the similarities in the general case are too striking to be accidental, and a deeper explanation of the connection with $\TT$ is definitely needed. 

The other issue with the ALD holographic correspondence that needs better explanation is more a general issue with the $\TT$ deformation. Namely, understanding what constitute good physical observables in such theories. For the class of examples we present here, some of these issues can be addressed. For example, perturbative open string theory provides a definition of a class of physical observables for the NCOS theory. 

We can hope that a deeper investigation of all these models with null boundaries will lead to a better understanding of what is needed for celestial or flat space holography~\cite{He:2014laa,Strominger:2017zoo,Raclariu:2021zjz,Pasterski:2021rjz,Donnay:2023mrd}. There are existing top down constructions of celestial holography~\cite{Costello:2022jpg,Costello:2023hmi,Bittleston:2024efo}, whose relation with the ALD models and the models presented here will be interesting to explore. Another approach to $D$-dimensional flat space holography involves a dual which is a $(D-1)$-dimensional Carrollian CFT; see, for instance, \cite{Bagchi:2023cen,Donnay:2022wvx,Bagchi:2025vri}.

At least in the NS5-F1 and the D1-D5-F1 examples coming from string theory, a non-local holographic theory with a Hagedorn density of states seems important for the existence of a null boundary. For any of these models, it will be interesting to extract a celestial theory, which determines the S-matrix, from the full holographic description. Some recent work looks particularly germane for this goal~\cite{Kervyn:2025wsb}.

\subsubsection*{\ul{\it A partial summary of results}}

The bulk of our paper contains a detailed step by step analysis but here we want to summarize the most important features.
We are considering boundstates of $n_1$ D1-branes, $n_5$ D5-branes and $m_1$ F1-strings of type IIB string theory in the NCOS decoupling limit. The parameters $(n_1, n_5, m_1)$ are integers. The D5-branes are wrapped on a $T^4$ of volume $16 \pi^4 \alpha^{\prime 2} \VT$ where $\frac{1}{2\pi \alpha'}$ is the fundamental string tension.   For the extremal case, the string-frame metric (\ref{extremalmetgoodcoord}) and  dilaton \C{extremaldilgoodcoord} are given in terms of dimensionless coordinates $({\tilde t}, {\tilde x}_5, u)$, which cover the $\AdS_3$ region at small $u$, the $S^3$ coordinates $(\theta, \phi, \psi)$ and the $T^4$ coordinates $( y_1, y_2, y_3, y_4)$:
\begin{empheq}[box=\doublebox]{align}
    \frac{ds^2_{\text{string}}}{\alpha'} &= \frac{m_1 \tilde{g}^2}{\VT} \frac{\sqrt{ f_1 f_5 } }{f_1 f_5 - 1 } \left( - d{\tilde t}^2 + d\tilde{x}_5^2 \right)  + \sqrt{ f_1 f_5 } \left( du^2 + u^2 \, d \Omega_3^2 \right) + \sqrt{ \frac{f_1}{f_5 } } \sqrt{\VT} \, {\widetilde ds}^2_{T^4} \, , \nonumber \\
    e^{2 \Phi} &=  \frac{ \tilde{g}^2 f_1^2 }{f_1 f_5 - 1 } \, . \nonumber
\end{empheq}
The coordinate ${\tilde x}_5 \sim {\tilde x}_5 + 2\pi {\tilde R}$ and $y_i \sim y_i + 2\pi$. The radial functions and $\tilde g$ are given by,
\begin{empheq}[box=\doublebox]{gather}
    f_1 = 1 + \frac{r_1^2}{u^2} \, , \qquad f_5 = 1 + \frac{r_5^2}{u^2} \, , \qquad r_1^2 = n_1  \frac{\tilde{g}}{\VT } \, , \qquad r_5^2 = n_5 \tilde{g} \, , \nonumber \\
    {\tilde g} = \frac{n_1}{m_1} + \frac{  n_5\VT}{m_1} \, . \nonumber
\end{empheq}
This metric interpolates between $M=0$ BTZ at small $u$ to an asymptotic Einstein-frame metric \C{nonextremalasymmetric} and dilaton \C{nonextremalasymdilaton}:
\begin{empheq}[box=\doublebox]{align}
    \frac{1}{\alpha'} ds^2_E &=  d\rho^2 +\frac{9}{16} \rho^2 \left( - d \tilde{t}^2 + d \tilde{x}_5^2 \right) + \frac{9}{16}  \rho^2\, d \Omega_3^2 + \ldots \, , \nonumber \\
    \Phi &= \frac{4}{3} \log ( \rho ) - \frac{2}{3} \log \left( \frac{16 m_1}{9 \VT}  \right) + \ldots \, , \nonumber
\end{empheq}
where $\rho$ is an asymptotic radial coordinate with large $\rho$ corresponding to large $u$, and omitted terms are suppressed by $\rho$. This is a six-dimensional spacetime.  There are associated NS and RR-fluxes described in section \ref{solutionsfromTST}. Features of the asymptotics and casual structure are described in sections \ref{sec:asymptotics} and \ref{sec:causal}. The asymptotic analysis of a scalar field propagating in this metric is also found in section \ref{sec:causal}. 

The RR scalar, $C_0$, has a non-trivial radial profile along with the dilaton. Notice that the string coupling diverges near infinity ($e^{2\Phi} \sim u^2 $) so one should S-dualize in that region to use perturbative string theory. Past gravity solutions that require different duality frames as one moves radially are described in~\cite{Martinec:1999bf}. The S-dual string frame metric, appropriate for the large $u$ region, is given in terms of the type IIB string coupling $\tau = C_0 + i e^{-\Phi}$ by $ds^2_{\rm S-dual} = |\tau| ds^2_{\rm string}$ where the S-dual dilaton is $e^{\hat \Phi} = |\tau|^2 e^{\Phi}$.\footnote{One might worry that the usual S transformation sending $\tau \rightarrow - \frac{1}{\tau}$ might not take us to weak string coupling because the $C_0$ potential also varies. Fortunately from \C{C0_soln} we see that $C_0$ decays at large $u$  and the usual S transformation still takes us to weak coupling. } In the sense of requiring patching by S-duality these backgrounds are S-folds, but the gravitational physics can be explored just using the Einstein-frame metric.

The generalization of these results to the non-extremal case is non-trivial and found in section \ref{sec:non_extremal}. We will just quote the metric \C{nonextremaldecoupled}, dilaton \C{nonextremaldilaton} and solution for $\tilde{g}$ \C{solutiong}:
\begin{empheq}[box=\doublebox]{align}
    \frac{ds^2}{\alpha'} &= \frac{m_1 \tilde{g}^2}{\VT}\frac{\sqrt{ f_1 f_5 } }{f_1 f_5 - f_e } \left( - f_e d{\tilde t}^2 + d \tilde{x}_5^2 \right)  + \sqrt{ f_1 f_5 } \left( \frac{1}{f_e} du^2 + u^2 \, d \Omega_3^2 \right) + \sqrt{ \frac{f_1}{f_5 } } \sqrt{\VT} \, {\widetilde ds}^2_{T^4}\, , \nonumber \\
    e^{2 \Phi} &= \frac{{\tilde g}^2 f_1^2 }{ f_1 f_5 -  f_e } \, , \qquad f_e = 1-\frac{r_e^2}{u^2}\,, \qquad  m_1^2 \tilde{g}^2 = n_1^2 + \VT \left( n_5^2 \VT + \sqrt{ 4 n_1^2 n_5^2 + m_1^2 r_e^4 } \right) \,.  \nonumber
\end{empheq}
It is important to note that $(r_1,r_5)$ now depend on $r_e$ via \C{constraints}. For positive $r_e^2>0$, there is a horizon at $u=r_e$ but there are also solutions with $r_e^2<0$. 
Note the density of states for large black holes has a Hagedorn growth described in section \ref{sec:mass}. The holographic dual description is an NCOS theory living on the D1-D5 brane system with parameters derived in section \ref{sec:ncos_holo}, 
\begin{empheq}[box=\doublebox]{gather}
    G_o^2 = \frac{n_1}{m_1}\,, \qquad \VT=\frac{n_1}{n_5} \, , \nonumber
\end{empheq}
where $G_o$ is the open string coupling and $\VT$ is fixed in the open string theory. Fixing $(n_1, n_5)$, we see that the open string theory has a good perturbative expansion for $m_1 \rightarrow\infty$.

There are many basic issues to be understood about this holographic correspondence from the way quantum information is encoded in the boundary theory to the classical data required for a well-posed initial value problem. Questions about classical field theory on these spacetimes to the structure, dynamics and stability of black holes to the theory of the S-matrix all call for a deeper investigation.

\vskip 0.2 in
\noindent {\bf Note added:} During the completion of this project, a paper appeared~\cite{Georgescu:2024iam} which examines aspects of related systems.

\section{Extremal Case} \label{sec:extremal}

We will be able to extract a great deal of data about this holographic correspondence by studying the extremal solution. The first step is constructing the extremal solution for the NCOS theory living on a collection of coincident $\hn_1$ D1-branes and $\hn_5$ D5-branes. We will not assume that the initial values of $(\hn_1, \hn_5)$ are integers. Rather we will impose flux quantization on the supergravity solution we obtain after applying the TsT duality chain and decoupling.

\subsection{Generating Solutions via TsT} \label{solutionsfromTST}

The usual procedure to generate a supergravity solution with a non-zero $B_2$-field employs the following steps~\cite{Maldacena:1999mh, Hashimoto:1999ut}. Start with a metric of the form, 
\begin{align}
    \label{general_metric}
    ds^2 &= h_1 ( r ) \, dx_0^2 + 2 h_2 ( r ) \, dx_0 \, dx_5 + h_3 ( r ) \, dx_5^2 + ds_{\perp}^2 \, ,  
\end{align}
where $r$ is a radial coordinate not involving $(x_0, x_5)$. 
We assume $x_5 \sim x_5 + 2\pi R$ and no initial $B_2$-field. The initial dilaton, $\Phi_0$, is assumed to be independent of $(x_0, x_5)$. Applying T-duality along $x_5$ using the standard rules of Appendix \ref{app:buscher} results in the supergravity background, 
\begin{align}
    ds^2 &= \left( h_1 ( r ) - \frac{h_2 ( r )^2}{h_3 ( r ) } \right) \, dx_0^2 + \frac{1}{h_3 ( r ) } \, dx_5^2 + ds^2_{\perp} \, , \nonumber \\
    B_{05} &= \frac{h_2 ( r ) }{h_3 ( r ) } \, , \qquad
    e^{2 \Phi} = \frac{1}{h_3} e^{2 \Phi_0} \, .
\end{align}
There are accompanying transformations of the Ramond fluxes also found in  Appendix \ref{app:buscher}. 

We now redefine coordinates from $(x_0, x_5)$ to $(x_0', x_5')$ using,
\begin{align}
    x_0 = x_0' \cosh ( \alpha ) + x_5' \sinh ( \alpha ) \, , \qquad x_5 = x_5' \cosh ( \alpha ) + x_0' \sinh ( \alpha ) \, ,
\end{align}
and express the metric in these coordinates,
\begin{align}
    ds^2 &= \left( \left( h_1 - \frac{h_2^2}{h_3} \right) \cosh^2 ( \alpha ) + \frac{\sinh^2 ( \alpha )}{h_3} \right) dx_0^{\prime 2} + 2 \cdot \frac{ \left( 1 - h_2^2 + h_1 h_3 \right) \cosh ( \alpha ) \sinh ( \alpha ) }{h_3} \, dx_0' \, dx_5'  \, \nonumber \\
    &\quad  + \left( \frac{\cosh^2 ( \alpha )}{h_3} + \left( h_1 - \frac{h_2^2}{h_3} \right) \sinh^2 ( \alpha ) \right) \, dx_5^{\prime 2} + ds_{\perp}^2 \, .
\end{align}
Lastly, we perform another T-duality along the $x_5$ direction to find
\begin{align}
    ds^2 &= \left( \frac{h_1}{\cosh^2 ( \alpha ) + \sinh^2 ( \alpha ) ( h_1 h_3 - h_2^2 ) } \right) \, dx_0^{\prime 2} + 2 \cdot \frac{h_2}{\cosh^2 ( \alpha ) + \sinh^2 ( \alpha ) \left( h_1 h_3 - h_2^2 \right) } \, dx_0^\prime \, dx_5^\prime \nonumber \\
    &\quad  + \frac{h_3}{\cosh^2 ( \alpha ) + \left( h_1 h_3 - h_2^2 \right) \sinh^2 ( \alpha ) } \, dx_5^{\prime 2} + ds_{\perp}^2 \, ,
\end{align}
along with
\begin{align}
    B_{05} = \frac{\cosh ( \alpha ) \sinh ( \alpha ) \left( 1 + h_1 h_3 - h_2^2 \right) }{\cosh^2 ( \alpha ) + \left( h_1 h_3 - h_2^2 \right) \sinh^2 ( \alpha ) } \, , \\ e^{2 \Phi} = \frac{e^{2 \Phi_0 ( r ) } }{ \cosh^2 ( \alpha ) + \sinh^2 ( \alpha ) \left( h_1 h_3 - h_2^2 \right) } \, .
\end{align}
This procedure produces the desired background with a $B_{05}$ potential.

\subsubsection*{\ul{\it Starting with the extremal D1-D5 solution}}

Let us apply the preceding TsT procedure to the extremal D1-D5 brane solution where the D5-branes are wrapping a four torus $T^4$ of volume $V_4$,
\begin{align}
    ds^2 = \frac{1}{\sqrt{f_1 f_5}} \left( - dt^2 + d x_5^2 \right) + \sqrt{ f_1 f_5 } \left( dr^2 + r^2 \, d \Omega_3^2 \right) + \sqrt{ \frac{f_1}{f_5 } } ds^2_{T^4} \, .
\end{align}
Here we have defined:
\begin{align}
    f_1 = 1 + \frac{\alpha' r_1^2}{r^2} = 1 + \frac{16 \pi^4 \alpha^{\prime 3} g \hn_1}{V_4 \, r^2} \, , \qquad f_5 = 1 + \frac{\alpha' r_5^2}{r^2} = 1 + \frac{\alpha' g \hn_5}{r^2} \, .
\end{align}
We take $x_5 \sim x_5 + 2 \pi R$ and choose coordinates $(\theta, \phi , \psi)$ for $S^3$. The initial dilaton and RR-fluxes take the form, 
\begin{align}
    e^{2 \Phi} &= \frac{g^2 f_1}{f_5 } \, , \nonumber \\
    F_3 &= - \frac{32 \alpha^{\prime 3} \pi^4 \hn_1}{V_{4} r^3 f_1^2} \, dt \wedge dr \wedge d x_5 + 2 \alpha' \hn_5 \sin^2 ( \theta ) \sin ( \phi ) \, d \theta \wedge d \phi \wedge d \psi \, .
\end{align}
The asymptotic value of the dilaton determines the string coupling, $g$.

After performing a TsT with parameter $\alpha$ along $x_5$, we generate a background with string-frame metric
\begin{align}\label{tststringframe}
    ds^2 = \frac{\sqrt{ f_1 f_5 } }{\cosh^2 ( \alpha ) f_1 f_5 - \sinh^2 ( \alpha ) } \left( - dt^2 + d x_5^2 \right) + \sqrt{ f_1 f_5 } \left( dr^2 + r^2 \, d \Omega_3^2 \right) + \sqrt{ \frac{f_1}{f_5 } } ds^2_{T^4} \, .
\end{align} 
The dilaton
\begin{align}\label{tstdilaton}
    e^{2 \Phi} = \frac{g^2 f_1^2}{\cosh^2 ( \alpha ) f_1 f_5 - \sinh^2 ( \alpha ) } \, , 
\end{align}
tends to $g^2$ as $r \to \infty$ since $f_i \to 1$. The NS flux $B_2$ is given by
\begin{align}
    B_2 &= \frac{  \sinh ( \alpha ) \cosh ( \alpha ) \left( f_1 f_5 - 1 \right) }{ \cosh^2 ( \alpha ) f_1 f_5 -  \sinh^2 ( \alpha ) } \, \, dt \wedge dx_5 \, ,
\end{align}
and the associated field strength is given by
\begin{align}
    H_3 = - \frac{ \sinh ( \alpha ) \cosh ( \alpha ) \left( f_5 ( r ) f_1 ' ( r ) + f_1 ( r ) f_5' ( r ) \right) }{\left( \cosh^2 ( \alpha ) f_1 ( r ) f_5 ( r ) - \sinh^2 ( \alpha ) \right)^2} \, dt \wedge dr \wedge d x_5 \, .
\end{align}
Lastly, we need to specify the resulting RR-fluxes starting with $F_3$ which takes the form
\begin{align}
    F_3 &= - \frac{32 \alpha^{\prime 3} \hn_1 \pi^4 \cosh ( \alpha ) f_5}{V_4 r^3 f_1 \left( \cosh^2 ( \alpha ) f_1 f_5 - \sinh^2 ( \alpha ) \right) } \, dt \wedge dr \wedge d x_5 \nonumber \\
    &\quad + 2 \cosh ( \alpha ) \alpha' \hn_5 \sin^2 ( \theta ) \sin ( \phi ) \, d \theta \wedge d \phi \wedge d \psi \, .
\end{align}
The post-TsT Ramond flux $F_1 = d C_0$ is
\begin{align}
    F_1 = - \frac{2  \alpha'  r_1^2 \sinh ( \alpha )}{g r^3 f_1^2} \, dr \, .
\end{align}
Likewise the self-dual flux $F_5$ has components
\begin{align}
    F_5 &= - \frac{2 \alpha' r_5^2 \sin^2 ( \theta ) \sin ( \phi ) \sinh ( \alpha )}{g \cosh^2 ( \alpha ) f_1 f_5 - g \sinh^2 ( \alpha ) } \, dt \wedge d \theta \wedge d \phi \wedge d \psi \wedge d x_5 \nonumber \\
    &\quad + \frac{2 \alpha' r_5^2 \sinh ( \alpha )}{g r^3 f_5^2 } \, dr \wedge dy_1 \wedge d y_2 \wedge d y_3 \wedge d y_4 \, , 
\end{align}
where the $y^i$ label the coordinates along the $T^4$.

\subsubsection*{\ul{\it NCOS Decoupling}}

Now we need to define the NCOS decoupling limit. With $\alpha' = \ell_s^2$ we follow the prior proposals and define, 
\begin{align}
    r = u \ell_s\, , \quad g = {\tilde g} \frac{b}{\alpha'} \, , \quad \cosh (\alpha) = \frac{b}{\alpha'}\, , \quad {\tilde x^i} = \frac{\ell_s}{b} x^i \, , \quad i=0, 5\, .
\end{align}
We will take the limit $\ell_s \rightarrow 0$ holding fixed $({\tilde g}, b, {\tilde x^i}) $. The resulting background is a solution of supergravity but is not, apriori, a solution of string theory. To be a solution of string theory, we need to impose flux quantization. 
Imposing quantization of $F_3$ through $S^3$ gives
\begin{align}
   \hn_5 = \frac{\alpha'}{b} n_5, \quad n_5 \in \Z\, .
\end{align}
This gives a finite expression, 
\begin{align} f_5 = 1 + \frac{r_5^2}{u^2}= 1 + \frac{n_5 \tilde{g}}{u^2}\,, \qquad r_5^2 = n_5 \tilde{g}  \, . 
\end{align}
Quantization of $F_7$ gives the relation ${\hat n}_1 = \frac{\alpha' }{b} n_1$. In terms of $n_1$, 
\begin{align}
    r_1^2 = n_1 \frac{16 \pi^4 g \alpha^{\prime 2}}{V_{4} \cosh ( \alpha ) }  = n_1  \frac{16 \pi^4 \tilde{g}  \alpha^{\prime 2}}{V_{4} } =  n_1  \frac{\tilde{g}}{\VT }  \, ,
\end{align}
where $n_1 \in \Z$. Here we have defined a finite torus volume ${\VT} = \frac{V_{4}}{16 \pi^4 (\alpha')^2}$ measured in string units and
\begin{align}
    f_1 =1 + \frac{ r_1^2}{{ u^2}} = 1 + \frac{ n_1 {\tilde g}}{{\VT u^2}} \, .
\end{align}
In this decoupling limit the dilaton becomes, 
\begin{align}\label{decoupled_dilaton}
    e^{2 \Phi} &= \frac{\tilde{g}^2 b^2}{\left( \alpha' \right)^2 } \cdot \frac{f_1^2}{1 + \cosh^2 ( \alpha ) \left( f_1 f_5 - 1 \right) } 
    = \tilde{g}^2 b^2 \cdot \frac{f_1^2}{\alpha^{\prime 2}  + b^2 \left( f_1 f_5 - 1 \right) } \nonumber \\
    &\rightarrow \frac{ \tilde{g}^2 f_1^2 }{f_1 f_5 -1 } =  \frac{ \tilde{g} f_1^2 \VT u^4 }{  n_1 n_5 {\tilde g} + \left( n_1   + n_5 \VT \right) u^2 } \, .
\end{align}
In the limit $u \rightarrow 0$, we see that 
\begin{align}
    e^{2 \Phi} \rightarrow \frac{ {\tilde g}^2  }{\VT}\frac{n_1}{n_5} \,,
\end{align}
which is the expected value in the $\mathrm{AdS}_3$ regime. In the large $u$ limit, we see that 
\begin{align}\label{largeudilaton}
    e^{2 \Phi} \rightarrow \frac{\tilde{g} \VT}{\left( n_1   + n_5 \VT\right)} u^2 \, ,
\end{align}
which diverges. The useful string description at large $u$ is given by the S-dual description which is a sector of the little string theory supported on NS5-branes. 

In the decoupling limit, the metric takes the nice form, 
\begin{align}\label{decoupled_metric}
    \frac{ds^2}{\alpha'} &= \frac{\sqrt{ f_1 f_5 } }{f_1 f_5 - 1 } \left( - d{\tilde t}^2 + d\tilde{x}_5^2 \right)  \nonumber \\
    &\quad + \sqrt{ f_1 f_5 } \left( du^2 + u^2 \, d \Omega_3^2 \right) + \sqrt{ \frac{f_1}{f_5 } } \sqrt{\VT} \, {\widetilde ds}^2_{T^4} \, .
\end{align}
Here $\widetilde{ ds}_{T^4}^2$ is the dimensionless metric for a torus with volume $(2 \pi)^4$.
Applying the decoupling limit to the fluxes gives, 
\begin{align}
    F_1 &= - \frac{2 n_1}{\VT f_1^2 u^3} du \, , \\
    F_3 &= \frac{2 n_1 f_5 \alpha' }{u^3 f_1 \left( f_1 f_5 -1 \right) \VT }  \, d {\tilde t} \wedge du \wedge d{\tilde  x}_5 \nonumber \\
    &\quad + 2 n_5 \alpha' \sin^2 ( \theta ) \sin ( \phi ) \, d \theta \wedge d \phi \wedge d \psi \,   \\
    F_5 &= - \frac{2 n_5 \alpha^{\prime 2}}{f_1 f_5 - 1} \, d {\tilde t} \wedge \left( \sin^2 ( \theta ) \sin ( \phi ) d \theta \wedge d \phi \right) \wedge d \psi \wedge d {\tilde x}_5 \nonumber \\
    &\quad + \frac{2 n_5 \alpha^{\prime 2}}{u^3 f_5^2 } \, du \wedge \tilde{\e}_4^{T^4} \\
    H_3 &= - \alpha' \frac{ \left( f_5 ( u ) f_1 ' ( u ) + f_1 ( u ) f_5' ( u ) \right) }{\left(  f_1 f_5  - 1 \right)^2} \, {d \tilde{t}} \wedge du \wedge {d \tilde{x}}_5 \, \\
    B_2 & = - \alpha' \left( \frac{1}{f_1 f_5 -1} \right)  \, {d\tilde{t}} \wedge {d \tilde{x}}_5 \, .
\end{align}
Note the dimensionless volume form for $T^4$, denoted $\tilde{\e}_4^{T^4}$ above, includes a factor of $\VT $, otherwise $F_5$ would not be self-dual. 
We have checked explicitly that this combination of metric, fluxes and dilaton solve the type IIB SUGRA equations of motion and the Bianchi identities.

\subsubsection*{\ul{\it Imposing string charge quantization}}

Our decoupled extremal solution does not appear to depend on $b$ but does depend on ${\tilde g}$. We now want to impose fundamental string charge quantization \C{H7quant} on the decoupled solution. There is still a transverse $S^3 \times T^4$ so we must still quantize this charge.  To impose quantization we need to evaluate the components of $\omega_7$, defined in \C{defomega7}, that lie along $S^3 \times T^4$. For this background, $\omega_7$ is given by the following expression:
\begin{align}
     -  \left( C_0 F_7 + C_4 \wedge F_3\right) \, .
\end{align}
We need the germane RR potentials which are easily obtained,
\begin{align}\label{C0_soln}
    & C_0(u) = \chi_0 +  \frac{n_1 }{\VT (u^2 + r_1^2)}  \, ,\\
    & C_4(u) = \chi_4 - \frac{ n_5 \alpha^{\prime 2}}{(u^2 + r_5^2)}\, ,  
\end{align}
where we have allowed $2$ constants $\chi_0$ and $\chi_4$ in the potentials.
We then find that the quantized charge is
\begin{align}
    m_1 = \frac{n_1}{\tilde{g}} + \frac{n_5 \VT}{ \tilde{g}} + n_1 \chi_0 - \frac{n_5 \chi_4}{ \alpha^{\prime 2}} \, .
\end{align}
The combination, 
\begin{align}
     n_1 \chi_0 = \frac{n_5 \chi_4}{\alpha^{\prime 2}}  \,, 
\end{align}
is massless while the orthogonal combination is massed up; see, for example,~\cite{Larsen:1999uk,Avery:2010qw}. Quantization of string charge then determines ${\tilde g}$ as follows:
\begin{align}
    {\tilde g} = \frac{n_1}{m_1} + \frac{  n_5\VT}{m_1} \, .
\end{align}
This agrees with \cite{Klebanov:2000pp} for the case $n_5=0$.

\subsection{Asymptotics}\label{sec:asymptotics}

The extremal solution is sufficient for us to determine the asymptotics that should define this quantum gravity theory. We will analyze the background in Einstein frame using the ten-dimensional relation, 
\begin{align}
  ds^2_{\rm Einstein} = e^{-\frac{\Phi}{2}} ds^2_{\rm string} \,.
\end{align}

\subsubsection*{\ul{\it Crossover behavior}}

The metric \C{decoupled_metric} studied in Einstein frame becomes $\AdS_3$ for very small $u$. We have been coy about specifying the periodicity of ${\tilde x}_5$. We have solutions of supergravity so we just declare that 
\begin{align}
    {\tilde x}_5 \sim {\tilde x}_5 + 2\pi {\tilde R}\,.
\end{align}
One might wonder why not set $\tilde R=1$? In the metric \C{decoupled_metric}, unlike $\AdS_3$ spacetime, the physical size of the circle stays finite as $u\rightarrow \infty$ so we want to retain the freedom of specifying that size via $\tilde{R}$.  We will use hatted coordinates to put the metric in canonical $\AdS$ form,  
\begin{align}
     &\frac{\sqrt{\tilde{g}} }{\alpha'} ds^2= \sqrt{r_1 r_5^3}\left[ \hat{u}^2 \left( - d{\hat t}^2 + d\hat{\varphi}^2 \right)  + \frac{1}{\hat u^2} \left( d{\hat u}^2 + {\hat u}^2 \, d \Omega_3^2 \right) \right]+ \sqrt{ \frac{r_1 \VT}{r_5}} \, {\widetilde ds}^2_{T^4} \, , \\ &\hat{\varphi} \sim \hat{\varphi} + 2\pi\,, \quad \hat{\varphi} = \frac{ {\tilde x_5}}{\tilde R}\,,\quad \hat{t}  = \frac{ \tilde{t}}{\tilde R}\,, \quad {\hat u} = \frac{\tilde R}{r_1 r_5 }u\,, \label{hatted}
\end{align}
with constant string coupling $e^{\Phi} = \tilde{g}\,\frac{r_1}{r_5} \sim \tilde{g} \sqrt{\frac{n_1}{n_5}}$. The $\AdS$ length scale in Einstein frame is $\ell^2 = \alpha' \sqrt{\frac{r_1 r_5^3}{{\tilde g}}}$. If we work in string-frame, the length scale would look more familiar $\ell^2_{\rm string} = \alpha' r_1 r_5 \sim \alpha' \sqrt{n_1 n_5}$.

This metric has three regimes: the very small $u$ $\AdS_3$ regime, the regime where the `$1$' in either $f_1$ or $f_5$ becomes non-negligible and the asymptotic regime where the `$1$' in both $f_1$ and $f_5$ is non-negligible. 
The `$1$'s in $f_1$ and $f_5$ become non-negligible when
\begin{align}
    u^2 \sim r_1^2 = \frac{ n_1 \tilde{g}}{\VT}, \qquad u^2 \sim r_5^2 = n_5 {\tilde g}\, ,
\end{align}
respectively. Which crossover regime we reach first depends on the choice of parameters $(n_1, n_5, \VT)$. For this extremal solution, $b$ makes no appearance. This is very similar to the extremal case of a partially decoupled NS5-F1 system; see~\cite{Chang:2023kkq} for a discussion. Instead of studying the crossover regimes in more detail, we will turn to the large $u$ asymptotics which should define the quantum gravity theory.

\subsubsection*{\ul{\it Large $u$ asymptotics}}

At large $u$, the Einstein frame metric takes the form
\begin{align} \label{largeuasymptotics}
    \frac{ds^2_{\rm Einstein}}{\alpha'} &=  A_1 u^{3/2} \left( 1+ O \left( \frac{1}{u^2}  \right) \right) \left( - d{\tilde t}^2 + d\tilde{x}_5^2 \right)  \nonumber \\
    &\quad + \frac{A_2}{\sqrt{u}} \left( 1+ O \left( \frac{1}{u^2}  \right)  \right) \left( du^2 + u^2 \, d \Omega_3^2 + \sqrt{\VT} {\widetilde ds}^2_{T^4} \right) \, .
\end{align}
where we have defined constants:
\begin{align}
    A_1 = \frac{m_1^{5/4} \VT^{3/4}}{( n_1 + \VT n_5)^2} \, , \qquad A_2 = \sqrt[4]{\frac{m_1}{\VT}} \, .
\end{align}
Staring at this metric, we see that the torus is small at large $u$ relative to the other factors but, more importantly, the $S^3$ factor is not negligible as $u \rightarrow \infty$. Going to a canonical radial variable $\rho = u^{3/4}$, we see at large $\rho$ the following asymptotic behavior
\begin{align} \label{asymptoticeinstein}
    \frac{ds^2_{\rm Einstein}}{\alpha'} &=  A_1\rho^2  \left( - d{\tilde t}^2 + d\tilde{x}_5^2 \right)  
   + A_2\left( \frac{16 } {9}  d\rho^2 + \rho^2 \, d \Omega_3^2 \right) + \ldots \, ,
\end{align} 
where omitted terms are subleading in $\rho$. Interestingly this asymptotic behavior matches the behavior seen for the partially decoupled NS5-F1 string system displayed in equation (A.28) of \cite{Chang:2023kkq}. Turning to the dilaton, we see that
\begin{align}\label{asymptoticdilaton}
    \Phi = \frac{4}{3} \log \rho + \frac{1}{2}\log\frac{\VT}{m_1} + O\left(\frac{1}{\rho^{8/3}}\right) \,.
\end{align}
This is also similar to partially decoupled NS5-F1 string system except the sign and coefficient of the leading log term are different.

Presumably this reflects the interesting spatial boundary structure which includes the $S^3$ factor appearing in \C{asymptoticeinstein}. So the boundary theory here really appears to be five-dimensional rather than the two-dimensional theory we might have expected from studying NCOS on D1-strings alone. In the case with no D5-branes, the NCOS metric can be found in~\cite{Harmark:2000wv}, for example. The asymptotic behavior of the metric in string-frame takes the approximate form:
\begin{align}
    ds^2_{\text{D1-branes}} \sim u^6\left( - d{\tilde t}^2 + d\tilde{x}_5^2 \right) + \left(  du^2 + u^2 \, d \Omega_7^2 \right) \, .
\end{align}
The boundary is two-dimensional in this case with no D5-branes. For our case, however, the D5-branes dominate the asymptotic behavior and our spatial boundary is conformally $\R \times S^1 \times S^3.$ 

\subsubsection*{\ul{\it The behavior of the curvature}}

One of the key characterizations of a quantum gravity theory is the behavior of the curvature as one approaches a boundary. For small $u$, we know the curvature will be negative and approximately constant since the solution looks like an $\AdS_3$ spacetime. What about for the large $u$ behavior of \C{largeuasymptotics}? Computing the Ricci scalar, one finds
\begin{align}
    \left(\alpha'\sqrt[4]{\frac{m_1}{\VT}}\right)  R = - \frac{ 1 }{2}  \frac{1}{u^{3/2}} - \frac{ n_5 \left( 5 n_1 - 18 n_5 \VT \right) }{8 m_1} \frac{1}{u^{7/2}} + O\left(\frac{1}{u^{11/2}}\right) \, ,
\end{align}
which approaches zero from below. 
The components of the Ricci tensor at large $u$ behave as
\begin{gather}
    R_{tt} = \frac{3 m_1 \VT}{2 ( n_1 + n_5 \VT )^2} - \frac{3 n_1 n_5 \VT}{2 ( n_1 + n_5 \VT )^2 u^2} + O \left( \frac{1}{u^4} \right) \, ,  \nonumber \\
    R_{x_5 x_5} = - \frac{3 m_1 \VT}{2 ( n_1 + n_5 \VT )^2} + \frac{3 n_1 n_5 \VT}{2 ( n_1 + n_5 \VT ) u^2} + O \left( \frac{1}{u^4} \right) \, , \qquad R_{uu} = - \frac{1}{u^2} + O \left( \frac{1}{u^4} \right) \, , \nonumber \\
    R_{\theta \theta} = \frac{1}{2} + O \left( \frac{1}{u^4} \right) \, , \qquad R_{\phi \phi} = \frac{1}{2} \sin^2 ( \theta ) + O \left( \frac{1}{u^4} \right) \, , \nonumber \\
    R_{\psi \psi} = \frac{1}{2} \sin^2 ( \theta ) \sin^2 ( \phi ) + O \left( \frac{1}{u^4} \right) \, ,  \qquad R_{y_i y_i} = \frac{ \sqrt{ \VT } }{2 u^2} + O \left( \frac{1}{u^4} \right) \, , 
\end{gather}
where $y_i$, $i = 1, 2, 3, 4$, label the $T^4$ directions. In particular, the Ricci scalar decays as $u^{-3/2}$ at large $u$, while the Ricci tensor has components which tend to constants at large $u$. The Einstein-frame solutions are therefore not asymptotically Ricci-flat.

\subsection{Causal Structure}\label{sec:causal}

One of the reasons this metric and its non-extremal generalizations are so interesting is that null geodesics take an infinite amount of time to reach the boundary. This is easy to check from \C{asymptoticeinstein}. Start at a fixed location on $\tilde{x}_5$ and the $S^3$ and fire a laser radially outward. It takes time, 
\begin{align}
    \Delta {\tilde{t}} \sim \int_{\rho=\rho_0}^{\rho_{\text{max}}} \frac{d\rho}{\rho} = \log \left( \frac{\rho_{\text{max}}}{\rho_0} \right) \, ,
\end{align}
to reach $\rho_{\text{max}}$, which diverges as $\rho_{\text{max}} \to \infty$. By contrast, photons from a laser fired in AdS will reach the boundary in finite time and bounce back. In our case, there is a genuine S-matrix for this holographic setup, which perhaps qualifies this correspondence as a top down example of celestial holography. 

\subsubsection*{\ul{\it Asymptotic scalar field propagation}}

Let us study an uncharged scalar field $\chi$ propagating in the asymptotic geometry described by (\ref{asymptoticeinstein}). We will explore the answer to two questions: how does the scalar field behave near spatial infinity? How does the scalar field behave near null infinity?

Let us first turn to the behavior near spatial infinity. The metric is sufficiently simple that the wave equation factorizes as follows:
\begin{align}\label{scalar_lap}
   \left[ \frac{9}{16 A_2} \p_\rho \left( \rho^5 \p_\rho \right) + \rho^3 \left( \frac{1}{A_1} \Box_{\tilde{t}\tilde{x_5}} +  \frac{1}{A_2} \Delta_{S^3}\right) \right] \chi =0\, .
\end{align}
The solution can be written in terms of eigenfunction of the operators in \C{scalar_lap} with coefficient $\rho^3$ of the form $e^{i\omega \tilde{t}} e^{i p_5 \tilde{x}_5} Y_\ell(\Omega_3)$ multiplied by a function of $\rho$ which we call $\chi_\rho$ so the total wavefunction takes the form $\chi = e^{i\omega \tilde{t}} e^{i p_5 \tilde{x}_5} Y_\ell(\Omega_3) \, \chi_\rho $. Defining the eigenvalue for the boundary wave operator of \C{scalar_lap}, 
\begin{align} \label{boundarymomentum}
    k^2 =  \frac{16}{9} \left( \frac{A_2}{A_1} \left( p_5^2 -\omega^2 \right) +\ell (\ell+2) \right) \, ,
\end{align}
leaves us with the radial equation, 
\begin{align}
    \left[  \p_\rho \left( \rho^5 \p_\rho  \right) - k^2 \rho^3  \right] \chi_\rho =0\, .
\end{align}
The general solution takes the form, 
\begin{align}\label{two_solns}
    \chi_\rho = a_1 \rho^{-2 + \sqrt{4 + k^2}} + a_2 \rho^{-2 - \sqrt{4 + k^2}} \, .
\end{align}
Ignoring the boundary coordinates, we can check normalizability at large $\rho$ with respect to the norm determined by the action, 
\begin{align}
 \int d\rho \, \rho^5 \left( \p_\rho \chi \p_\rho \chi + \frac{1}{\rho^2} \chi_\rho^2 \right) \,.
\end{align}
The second term above comes from kinetic terms like $g^{\tilde{x_5}\tilde{x_5}} \p_{\tilde{x_5}} \chi \p_{\tilde{x_5}} \chi$ which involve boundary derivatives, but we have omitted the boundary momentum factors which play no role in checking $\rho$ normalizability. From this we see that the $c_1$ solution of \C{two_solns} is not normalizable but the $c_2$ solution is normalizable as long as $k^2 > -4$. This is similar in spirit to the analysis of a scalar in AdS except the resulting asymptotic behavior seen in \C{two_solns} depends on the boundary momentum. 

Adding a mass term modifies the equation as follows, 
\begin{align}
   \left[  \p_\rho \left( \rho^5 \p_\rho  \right) - k^2 \rho^3 - m^2 \rho^5  \right] \chi_\rho =0\, ,
\end{align}
and the corresponding solution is given in terms of Bessel functions, 
\begin{align} \label{massivesoln}
    \chi_\rho = \frac{1}{\rho^2}\left(a_1 I_{\sqrt{4+k^2}}(m\rho) + a_2 K_{\sqrt{4+k^2}}(m\rho) \right)\,. 
\end{align}
At large $\rho$, the asymptotic behavior of \C{massivesoln} takes the form,
\begin{align} \label{massivexpansion}
    \chi_\rho \sim \frac{1}{\rho^2} \left(a_1 \frac{e^{m\rho}}{\sqrt{2\pi m\rho}} \left[ 1 - \frac{a_3}{m\rho}\right] + a_2   e^{-m\rho} \sqrt{\frac{\pi}{2 m\rho}} \left[ 1 + \frac{a_3}{m\rho}\right]\right)\,, \quad a_3 = \frac{15+4k^2}{8}\, ,
\end{align}
so the $a_1$ solution is not normalizable while the $a_2$ solution decays exponentially. For the massive scalar, the boundary momentum only appears at subleading order in \C{massivexpansion}.

\subsubsection*{\ul{\it Scalar field radiation}}

To repeat this analysis near null infinity for a massless scalar, we need to write the metric in Bondi-like coordinates, which are well adapted for radiation.  Let us define a $v$ variable in two steps. First define a new coordinate $v$ in terms of $\tilde{t}$ and $u$ so that the coordinate choice kills the $du^2$ term in the metric:
\begin{align}\label{fulldiff}
    \tilde{t} = v \pm \frac{ ( n_1 + n_5 \VT ) \left( \sqrt{ n_1 n_5 + m_1 u^2} + u \sqrt{m_1} \log \left( \sqrt{ n_1 n_5 + m_1 u^2 } - \sqrt{m_1} u \right) \right)}{ m_1 u \sqrt{ \VT } } \, .
\end{align}
We will study the minus case in \C{fulldiff} here which is suitable for describing outgoing waves. The second step is defining $\rho = u^{3/4}$ which is convenient for comparison with the preceding discussion. We will use the full diffeomorphism \C{fulldiff} in the analysis below but it is helpful to just see the leading relation, 
\begin{align}\label{leadingdiff}
    \tilde{t} = \frac{4}{3} \sqrt{ \frac{A_2}{A_1} } \log ( \rho ) + v \, ,
\end{align}
in terms of which the metric takes the form
\begin{align} \label{nullmetric}
    \frac{1}{\alpha'} ds^2_{(0)} = \rho^2 \left( - A_1 \, dv^2 + A_1  \, dx_5^2 + A_2  \, d \Omega_3^2 \right) - \frac{8}{3} \sqrt{A_1 A_2} \, \rho \, dv \, d \rho  \, ,
\end{align}
with no $d \rho^2$ term, so that $\rho$ is indeed a null coordinate. We will also want the first subleading metric corrections
\begin{align} \label{subleadingmetric}
    \frac{1}{\alpha'} ds_{(1)}^2 =  \frac{A_3}{\rho^{2/3}} \left( - dv^2 + dx_5^2 \right) - \frac{A_4}{\rho^{5/3}} \, dv \, d \rho
    + \frac{A_5}{\rho^{2/3}} d \Omega_3^2 + \frac{A_6}{\rho^{2/3}} \widetilde{ds}^2_{T^4}  + O \left( \frac{1}{\rho^{10/3}} \right)\,,
\end{align}
where the constants are listed below:
\begin{gather}
    A_3 = \frac{\sqrt[4]{m_1} \VT^{3/4} ( 2 n_5^2 \VT  - n_1 n_5 )}{ 4 ( n_5 \VT + n_1 )^2} \, , \quad A_4 = \frac{2 \sqrt[4]{\VT} ( 2 n_5^2 \VT + n_1 n_5 )}{3 \sqrt[4]{m_1} ( n_5 \VT + n_1) } \, , \nonumber \\
    A_5 = \frac{3 n_1 n_5 + 2 n_5^2 \VT}{4 m_1^{3/4} \sqrt[4]{\VT}} \, , \quad A_6  = \sqrt[4]{m_1 \VT} \,.
\end{gather}
In the following discussion we will ignore momentum along the torus directions because any particle carrying torus momentum will be massive and therefore cannot reach $\mathscr{I}^+$.

The wave equation for a massless scalar $\chi$ in the metric (\ref{nullmetric}) taking into account the relevant terms from \C{subleadingmetric} is
\begin{align} \label{leadingwave}
    & \Box_{(0)} \chi^{(0)} = - \frac{1}{16 A_1 A_2 \rho^2} \Big[  \sqrt{A_1 A_2} \left( 32 \frac{\partial}{\partial v} + 24\rho \frac{\partial^2}{\partial v \, \partial \rho} \right) -  A_1 \rho \left( 33 \frac{\partial}{\partial \rho} + 9 \rho \frac{\partial^2}{\partial \rho^2} \right) \nonumber \\
    &\qquad \qquad \qquad - 16 A_2 \frac{\partial^2}{\partial x_5^2} - 16A_1 \,\Delta_{S^3} \Big] \chi^{(0)} = 0 \, .
\end{align}
Similar to \C{boundarymomentum}, let us define a boundary momentum $k_B$ as follows
\begin{align}
    k_B^2 =  16  \left(  A_2 p_5^2 +A_1 \ell (\ell+2) \right) \, ,
\end{align}
and rewrite the wave equation \C{leadingwave}:
\begin{align} \label{leadingwave2}
    &\Big[  \sqrt{A_1 A_2} \left( 32 \frac{\partial}{\partial v} +  24 \frac{\partial^2}{\partial v \, \partial \log\rho} \right) -  A_1 \left( 24 \frac{\partial}{\partial \log \rho} +  9 \frac{\partial^2}{\partial (\log \rho)^2} \right)  + k_B^2 \Big] \chi^{(0)} = 0 \, .
\end{align}
The dependence of the leading order solution on $v$ and $\rho$ takes the form, 
\begin{align} \label{leadingscalar}
     \chi^{(0)} = e^{i E \left( v + \frac{4}{3}\sqrt{\frac{A_2}{A_1}} \log\rho \right) + k_\rho \log\rho}\,, \qquad E^2 = \frac{1}{16 A_2} \left( k_B^2 - 3 A_1 k_\rho ( 8 + 3 k_\rho ) \right) \,,
\end{align}
giving us the dispersion relation for the wave. The form of the solution $\chi^{(0)}$ is an oscillating plane wave multiplied by a radial function. For any given boundary momentum $k_B^2$, which is real, we must insist that 
\begin{align}
    3A_1 k_\rho (8+3 k_\rho) < k_B^2 \,, 
\end{align}
so that the energy $E$ is still real. When $k_\rho =0$ or $k_\rho = - \frac{8}{3}$, which is a decaying solution, the energy $E^2 = \frac{k_B^2}{16A_2}$. For a real $E$, the range of permitted values for $k_\rho$ is given by, 
\begin{align} \label{solutionkrho}
 - \frac{4}{3} - \sqrt{ 16+ \frac{k_B^2}{A_1}}    \leq k_\rho \leq - \frac{4}{3} + \sqrt{ 16+ \frac{k_B^2}{A_1}} \, .
\end{align}
If $k_\rho$ is positive definite then the solution is growing at large $\rho$. Decaying solutions require $k_\rho$ negative. For example, if we set the boundary momentum $k_B=0$ and demand a decaying solution then 
\begin{align}
    -\frac{16}{3} \leq k_\rho \leq 0\, .
\end{align}
If we further insist on a wavefunction with $L^2$ normalizable decay near null infinity with a norm determined by the leading metric \C{nullmetric} with determinant going like $\rho^5$ then we want $k_\rho < -2$.

We will close this discussion of scalar field radiation by sketching how the first subleading correction to \C{leadingscalar} emerges.  For this we need the subleading corrections to \C{leadingwave} which we denote $\Box_{(1)}$. To determine this operator, we start from the exact metric and then compute the Laplacian and expand in powers of $\rho$; simply using the subleading corrections to the metric encoded in \C{subleadingmetric} can miss relevant volume factors in the Laplacian. The leading corrections are down by $O(\frac{1}{\rho^{8/3}})$ from $\Box_{(0)}$.  We want to solve an equation of the form, 
\begin{align} \label{subleadingexpansion}
   \Box_{(0)} \chi^{(1)} = - \Box_{(1)} \chi^{(0)}\,,
\end{align}
for $\chi^{(1)}$. This process is analogous to the Fefferman-Graham expansion \cite{AST_1985__S131__95_0,2007arXiv0710.0919F} near the boundary of $\AdS$ but instead of an asymptotic expansion near spatial infinity, we are instead integrating inward from null infinity. It would be nice to explore the systematics of this expansion in more detail. Particularly the patching of the asymptotic solution to the wavefunction of a scalar in the $\AdS$ region where massless scalars behave as follows in terms of radial dependence:
\begin{align}
    \phi^{\AdS} \sim \phi^{(0)} + \frac{\phi^{(2)}}{r^2} \,.
    \end{align}
The term $\phi^{(2)}$ has non-zero $\AdS$ energy. The time scales governing the rate at which black holes decay via emission of scalar radiation should then have multiple regimes depending on the size of the black hole compared to the three crossover regimes. 

\subsection{A more convenient coordinate choice}

So far our discussion of the extremal solution has used the coordinate system that naturally come from the TsT procedure followed by decoupling. The solution was agnostic to the periodicity of ${\tilde x}_5$, which we simply declared to be $2\pi\tilde{R}$. 

In our later discussion, we will construct the non-extremal solutions. To fit the extremal solution smoothly into that family of solutions, it will be useful to change coordinates. We will redefine $\tilde{t}$ and $\tilde{x}_5$ so that the extremal metric and dilaton takes the form,
\begin{align} \label{extremalmetgoodcoord}
    \frac{ds^2}{\alpha'} &= \frac{m_1 \tilde{g}^2}{\VT} \frac{\sqrt{ f_1 f_5 } }{f_1 f_5 - 1 } \left( - d{\tilde t}^2 + d\tilde{x}_5^2 \right)   + \sqrt{ f_1 f_5 } \left( du^2 + u^2 \, d \Omega_3^2 \right) + \sqrt{ \frac{f_1}{f_5 } } \sqrt{\VT} \, {\widetilde ds}^2_{T^4} \, , \\ \label{extremaldilgoodcoord}
    e^{2 \Phi} &=  \frac{ \tilde{g}^2 f_1^2 }{f_1 f_5 - 1 } \, , \qquad {\tilde g} = \frac{n_1}{m_1} + \frac{  n_5\VT}{m_1} \, .
\end{align}
Under this diffeomorphism, the associated fluxes take the form:
\begin{align} \label{extremalfluxgoodcoord}
    H_3 &= - \alpha' \frac{m_1}{\VT} \tilde{g}^2 \frac{ \left( f_1 f_5' + f_5 f_1' \right) }{\left(  f_1 f_5 - 1  \right)^2 } \, d {\tilde{t}} \wedge du \wedge d {\tilde{x}}_5 \, ,  \qquad F_1 =  - \frac{2 n_1}{\VT u^3 f_1^2} du\,, \nonumber \\
    F_3 &= - \frac{2 \alpha' \tilde{g}^2 m_1 n_1}{\VT^2 u^3 f_1} \cdot  \frac{f_5 }{ f_1 f_5 - 1 } d\tilde{t}\wedge d\tilde{x}_5 \wedge du +  2 \alpha' n_5 \sin^2 ( \theta ) \sin ( \phi )  d \theta \wedge d \phi \wedge d \psi \, , \nonumber \\
    F_5 &= \frac{2 \tilde{g}^2 m_1  n_5 \alpha^{\prime 2}}{ \VT \left( f_1 f_5 - 1 \right) } d\tilde{t}\wedge d\tilde{x_5} \wedge \sin^2 ( \theta ) \sin ( \phi ) d \theta \wedge d \phi \wedge d \psi + \frac{2 n_5 \alpha^{\prime 2} \VT}{u^3 f_5^2} \, du \wedge \widetilde{\e}_4^{T^4} \, .
\end{align}
We will  declare that this $\tilde{x}_5$ is periodic with 
$\tilde{x}_5 \sim \tilde{x}_5 + 2\pi \tilde{R}$. 

\subsection{NCOS Holographic Dual}\label{sec:ncos_holo}

We now turn to the holographic definition of quantum gravity with the asymptotic behavior specified in \C{largeuasymptotics}. This is a theory with spacetime non-commutativity. 
To derive the holographic theory, one applies the decoupling limit to the system of D1-D5 branes with $B_2$ supported on the D1-string world-volume. Unlike the more familiar decoupling limits which result in $\AdS$ spacetimes, in this case the resulting theory on the brane system is not a local field theory but rather a non-commutative open string theory (NCOS). Specifically the open string oscillator modes do not decouple. This is very much what we might suspect is required for a spacetime with a null boundary. 

The other known example is linear dilaton holography where the holographic definition is a kind of $\TT$-deformed CFT. In that case, the precise definition of the holographic dual is not known but we do know that the high-energy density of states exhibits Hagedorn growth, which is also the case with our model at weak coupling. 

We want to decouple a collection of $n_1$ D1-branes and $n_5$ D5-branes together with $m_1$ F1-strings or units of electric flux. The NCOS decoupling limit requires taking the following scalings:
\begin{align} \label{branedecoupling}
    \alpha' = \alpha'_e \epsilon\,, \qquad g_s = \frac{G_o^2}{\sqrt{\epsilon}}\,, \qquad 2\pi\alpha' \e^{tx_5} F_{tx_5} = 2\pi \alpha' E =  1 - \frac{\epsilon}{2}\,,
\end{align}
with the electric field on the brane approaching the critical value.  We have defined the open string coupling $G_o^2$ in \C{branedecoupling}.   
In addition, we do not scale the metric in the $(t,x_5)$ directions which support the electric field but we do scale the metric in the remaining brane world-volume directions $g_{ij}$ as well as the metric for the directions transverse to the brane $g_{MN}$ with open string parameters
\begin{align}
   G_{\m\n} = \epsilon\eta_{\m\n}\,, \qquad  G_{ij} = \epsilon g_{ij}\,, \qquad G_{MN} = \epsilon g_{MN}\, ,
\end{align}
where we have assumed $g_{\m\n} = \eta_{\m\n}$ for simplicity.

First let us examine the gauge couplings for the D1-branes and the D5-branes, respectively. For this purpose, we can initially set $n_1=n_5=1$. For the D1-brane, we expand the DBI action
\begin{gather}
    S_{\text{DBI}} = - T_{\text{D}1} \int d^2 \sigma \, e^{- \Phi} \sqrt{ - \det ( \eta_{\mu \nu} + 2 \pi \alpha' F_{\mu \nu} ) } \,, \nonumber \\
    T_{\text{D}1} = \frac{1}{2 \pi \alpha' g_s} \, , \qquad \det ( \eta_{\mu \nu} + 2 \pi \alpha' F_{\mu \nu} ) = - 1 + 4 \pi^2 \alpha^{\prime 2} (E+\hat{F}_{01})^2 \, ,
\end{gather}
where we are only  focusing on the world-volume directions and fluctuations around the constant electric field are denoted by $\hat{F}$. Expanding and using \C{branedecoupling} gives,
\begin{align}
    S_{\text{DBI}} &= \frac{\pi \alpha'_e}{2 {G_o^2}} \int d^2 \sigma \, \sqrt{G} e^{- \Phi} \hat{F}_{\mu\nu} \hat{F}_{\mu'\nu'} G^{\mu\mu'} G^{\nu\nu'}  \, , 
\end{align}
with Yang-Mills gauge coupling $g^2_{(1)} = \frac{G_o^2}{2\pi \alpha'_e}$. A similar analysis for the D5-brane with tension $ T_{\text{D}5} = \frac{1}{(2 \pi)^5 (\alpha')^3 g_s}$ gives Yang-Mills coupling $g^2_{(5)} =  (2\pi)^3 \alpha'_e G_o^2$. Both couplings are finite, which is quite different from the decoupling limit that gives $\AdS_3$ where only the D1-string gauge coupling remains finite. The non-trivial dynamics on the D5-branes is the reason our spacetime boundary is five-dimensional and not two-dimensional.

What remains is to determine $G_o^2$ in terms of the flux quantum numbers. We have $m_1$ units of $B_{0x_5}$ corresponding to the dissolved fundamental strings. We gauge $B_{0x_5}$ into $F_{(D1)}$ and $F_{(D5)}$ and take the critical electric field limit. The electric displacements, which are the conjugate momenta to the $x_5$ components of $A_{(D1)}$ and $A_{(D5)}$,  are quantized on a compact space with
\begin{align}
   T_{D1} \frac{ ( 2 \pi \alpha' )^2 E }{\sqrt{ 1 - ( 2 \pi \alpha' )^2 E^2 } } = m_1\, , \qquad T_{D5} V_4 \frac{ ( 2 \pi \alpha' )^2 E }{\sqrt{ 1 - ( 2 \pi \alpha' )^2 E^2 } } = m_1\, ,
\end{align}
as discussed in~\cite{Gopakumar:2000na}. The first relation fixes $G_o^2 = \frac{1}{m_1}$. The second quantization condition then forces us to fix the volume $\VT=1$ at the location of the brane. Although there is no current closed form expression for the non-abelian brane effective action, we can extend the quantization condition to $n_1$ D1-branes and $n_5$ D5-branes by studying the Coulomb branch where the branes are slightly separated, giving
\begin{align}\label{holographic_params}
    G_o^2 = \frac{n_1}{m_1}\,, \qquad \VT=\frac{n_1}{n_5}\, ,
\end{align}
for the open string theory.

\section{Non-Extremal Case}\label{sec:non_extremal}

\subsection{Constructing the solutions}\label{sec:constructing_non_extremal}

Now we would like to extend our preceding discussion of the extremal solution to non-extremal black hole solutions. We will restrict to solutions without spin. There are interesting subtleties in applying the TsT procedure in the non-extremal case because of the existence of a horizon.\footnote{\label{troels_footnote} This issue was described in~\cite{Harmark:2000wv}. We would like to thank T. Harmark for helpful correspondence on this point in September of 2023.} We initially want to follow our nose and repeat the steps used to generate the NCOS extremal solution. 

\subsubsection*{\ul{\it Starting with the non-extremal D1-D5 solution}}

So we start with the non-extremal D1-D5 metric~\cite{David:2002wn,Shin:2007gz},
\begin{align}
    ds^2 = \frac{1}{\sqrt{f_1 f_5}} \left( - f_e \, dt^2 + dx^2_5 \right) + \sqrt{ f_1 f_5 } \left( \frac{dr^2}{f_e} + r^2 d \Omega_3^2 \right) + \sqrt{ \frac{f_1}{f_5}} ds^2_{T^4} \, ,
\end{align}
where
\begin{align}
    f_1 = 1 + \frac{\alpha' r_1^2}{r^2} \, , \qquad f_5 = 1 + \frac{\alpha' r_5^2}{r^2} \, , \qquad f_e = 1 - \frac{\alpha' r_e^2}{r^2} \, .
\end{align}
There is a horizon at $r=r_e$, which is the non-extremality parameter. The associated dilaton and $F_3$-field take the same form as the extremal case, 
\begin{align}
    & e^{2 \Phi} = g^2 \frac{f_1}{f_5} \, , \\
   & F_3 = 
   - \frac{32 \alpha^{\prime 3} \pi^4 \hn_1}{V_{4} r^3 f_1^2} \, dt \wedge dr \wedge d x_5 + 2 \alpha' \hn_5 \sin^2 ( \theta ) \sin ( \phi ) \, d \theta \wedge d \phi \wedge d \psi \, .
\end{align}
Applying the same procedure as in section \ref{sec:extremal} gives the dilaton,
\begin{align} \label{undecouplednonextremaldilaton}
    e^{2 \Phi} = \frac{g^2 f_1^2 }{\cosh^2 ( \alpha ) f_1 f_5 - \sinh^2 ( \alpha ) f_e } \, , 
\end{align}
which tends to $g^2$ as $r \to \infty$ since $f_i \to 1$. The metric is given by
\begin{align} \label{nonextremalundecoupled}
    ds^2 = & \frac{\sqrt{ f_1 f_5 }  }{\cosh^2 ( \alpha ) f_1 f_5 - f_e \sinh^2 ( \alpha ) } \left( - f_e \, dt^2 + d x_5^2 \right) + \sqrt{ f_1 f_5 } \left( \frac{dr^2}{f_e} + r^2 \, d \Omega_3^2 \right) \cr & + \sqrt{ \frac{f_1}{f_5 } } ds^2_{T^4} \, ,
\end{align}
while the $B_2$-field and associated $H_3$ take the form, 
\begin{align}
    B_{tx_5} &= \cosh(\alpha) \sinh(\alpha) \frac{f_1f_5- f_e}{f_1f_5\cosh^2(\alpha) - f_e \sinh^2(\alpha)} \, , \\
    H_{t u x_5} &= \frac{\cosh ( \alpha ) \sinh ( \alpha ) \left( f_1 f_5 f_e' - f_e \left( f_1 f_5' + f_5 f_1' \right) \right) }{\left( \cosh^2 ( \alpha ) f_1 f_5 - f_e \sinh^2 ( \alpha ) \right)^2 } \, . 
\end{align}
Finally the RR-fluxes are given by,
\begin{align}
    F_1 &=  - \frac{2   \alpha' \hn_1 }{\VT r^3 f_1^2} \sinh(\alpha) dr\,, \\
    F_3 &= - \frac{2   \alpha' \hn_1}{\VT r^3 f_1^2} \cosh(\alpha) \frac{ f_1 f_5 }{\cosh^2 ( \alpha ) f_1 f_5 - \sinh^2 ( \alpha ) f_e } dt\wedge dx_5 \wedge dr \nonumber \\
    &\quad +  2  \alpha' \hn_5  \cosh ( \alpha ) \sin^2 ( \theta ) \sin ( \phi )  d \theta \wedge d \phi \wedge d \psi \, , \\
    F_5 &= \frac{2 \hn_5 \alpha'  \sinh ( \alpha )  f_e}{\cosh^2 ( \alpha ) f_1 f_5 - \sinh^2 ( \alpha ) f_e} dt\wedge dx_5 \wedge \sin^2 ( \theta ) \sin ( \phi ) d \theta \wedge d \phi \wedge d \psi \nonumber \\
    &\quad + \frac{2 \hn_5 \alpha' \sinh ( \alpha ) }{r^3 f_5^2} \, dr \wedge d y^1 \wedge d y^2 \wedge dy^3 \wedge d y^4 \, .
\end{align}
Note that the pre-TsT values of $\hn_1$ and $\hn_5$ are {\it not} quantized. We will impose quantization on the post-TsT background. 

\subsubsection*{\ul{\it Decoupled non-extremal solutions}}

Now let us decouple using the NCOS decoupling limit. We define a dimensionless $u$ as before,
\begin{align}
    f_1 = 1 + \frac{r_1^2}{u^2} \, , \quad f_5 = 1 + \frac{r_5^2}{u^2} \, , \quad f_e = 1 - \frac{r_e^2}{u^2} \, .
\end{align}
For the moment, we will assume that $r_e^2 \geq 0$. The function $f_e$ only blows up at the origin like $f_1$ and $f_5$. On decoupling, we now encounter quantities that involve $ (\alpha')^2 f_e$ in the $\alpha' \rightarrow 0$ limit; for example in the dilaton
\begin{align} 
    e^{2 \Phi} = \frac{{\tilde g}^2 f_1^2 }{ f_1 f_5 -  f_e + \left(\frac{\alpha'}{b} \right)^2 f_e} \, .
\end{align}
We will drop $ (\alpha')^2 f_e$ term because the potentially problematic divergence is at the origin screened by a horizon. This is true for all cases aside from states with negative $r_e^2$ where the origin is potentially accessible. We will assume this term can still be dropped even in those cases. With this caveat the dilaton is straightforwardly determined giving, 
\begin{align} \label{nonextremaldilaton}
    e^{2 \Phi} = \frac{{\tilde g}^2 f_1^2 }{ f_1 f_5 -  f_e } \, . 
\end{align}
In the decoupling limit, the metric again takes the nice form, 
\begin{align} \label{nonextremaldecoupled}
    \frac{ds^2}{\alpha'} &= \frac{c_1\sqrt{ f_1 f_5 } }{f_1 f_5 - f_e } \left( - f_e d{\tilde t}^2 + d \tilde{x}_5^2 \right)  \nonumber \\
    &\quad + \sqrt{ f_1 f_5 } \left( \frac{1}{f_e} du^2 + u^2 \, d \Omega_3^2 \right) + \sqrt{ \frac{f_1}{f_5 } } \sqrt{\VT} \, {\widetilde ds}^2_{T^4} \, .
\end{align}
We are allowing a (yet) undetetermined constant $c_1$ for later convenience. 
The issue of mass-dependence of the asymptotics will need to be resolved later and this will determine $c_1$. This is the issue mentioned in footnote \ref{troels_footnote}. If we momentarily ignore the periodicity of $\tilde{x}_5$ then the choice of $c_1$ is a choice of definition of $(\tilde{t}, \tilde{x}_5)$. Finally the $H_3$-field takes the form, 
\begin{align}
    H_3 &= \alpha' c_1 \frac{ \left( f_1 f_5 f_e' - f_e \left( f_1 f_5' + f_5 f_1' \right) \right) }{\left(  f_1 f_5 - f_e  \right)^2 } \, d {\tilde{t}} \wedge du \wedge d {\tilde{x}}_5 \,  , \\
    B_2 & = - \alpha' c_1 \left( \frac{f_e}{f_1 f_5 -f_e} \right)  \, {d\tilde{t}} \wedge d {\tilde{x}}_5 \, .
\end{align}

The decoupled RR-fluxes take the form:
\begin{align}
    F_1 &=  - \frac{2 b \hn_1}{\alpha' \VT u^3 f_1^2} du\,, \\
    F_3 &= - \frac{2 b c_1 \hn_1}{\VT u^3 f_1} \cdot  \frac{f_5 }{ f_1 f_5 - f_e } d\tilde{t}\wedge d\tilde{x}_5 \wedge du \nonumber \\
    &\quad +  2 \hn_5  b \sin^2 ( \theta ) \sin ( \phi )  d \theta \wedge d \phi \wedge d \psi \, , \\
    F_5 &= \frac{2 \hn_5 b c_1 \alpha'   f_e}{ f_1 f_5 - 1 } d\tilde{t}\wedge d\tilde{x_5} \wedge \sin^2 ( \theta ) \sin ( \phi ) d \theta \wedge d \phi \wedge d \psi \nonumber \\
    &\quad + \frac{2 \hn_5 b \alpha' \VT}{u^3 f_5^2} \, du \wedge \widetilde{\e}_4^{T^4} \, .
\end{align}
The quantized quantities arising from integrating $F_3$ and its dual are
\begin{align}
    n_5 = \frac{b \hn_5}{\alpha'} \, , \qquad n_1 = \frac{b \hn_1}{\alpha'} \, .
\end{align}
Here are the final quantized fluxes:
\begin{align} \label{finalquantfluxes}
    F_1 &=  - \frac{2 n_1}{\VT u^3 f_1^2} du\,, \\
    F_3 &= - \frac{2 c_1 \alpha' n_1}{\VT u^3 f_1} \cdot  \frac{f_5 }{ f_1 f_5 - f_e } d\tilde{t}\wedge d\tilde{x}_5 \wedge du \nonumber \\
    &\quad +  2 \alpha' n_5 \sin^2 ( \theta ) \sin ( \phi )  d \theta \wedge d \phi \wedge d \psi \, , \\
    F_5 &= \frac{2 c_1 n_5 \alpha^{\prime 2}   f_e}{ f_1 f_5 - f_e } d\tilde{t}\wedge d\tilde{x_5} \wedge \sin^2 ( \theta ) \sin ( \phi ) d \theta \wedge d \phi \wedge d \psi \nonumber \\
    &\quad + \frac{2 n_5 \alpha^{\prime 2} \VT}{u^3 f_5^2} \, du \wedge \widetilde{\e}_4^{T^4} \, .
\end{align}
In order to solve the Einstein equations and the $B_2$-field equation of motion, we must impose the following algebraic constraints between the parameters:
\begin{align} \label{constraints}
    r_5^2 &= \frac{1}{2} \left( \sqrt{ 4 \tilde{g}^2 n_5^2 + r_e^4 } - r_e^2 \right) \, , \nonumber \\
    r_1^2 &= \frac{1}{2} \left( \sqrt{ \frac{4 \tilde{g}^2 n_1^2}{ \VT }  + r_e^4 } - r_e^2 \right) \, .
\end{align}
Note that $(r_1, r_5)$ are deformed away from their extremal values but still always positive. To complete the description of this background, we finally impose string charge quantization.
The quantized charge $m_1$ associated with $H_7$ is
\begin{align}
    m_1 = \frac{1}{2 \tilde{g}^2} \left( \sqrt{ 4 \tilde{g}^2 n_5^2 + r_e^4 } \VT + \sqrt{ 4 \tilde{g}^2 n_1^2 + r_e^4 \VT^2 } \right) \, ,
\end{align}
Solving for $\tilde{g}^2$ gives
%
\begin{align} \label{solutiong}
    \tilde{g}^2 = \frac{n_1^2}{m_1^2} + \frac{\VT}{m_1^2} \left( n_5^2 \VT + \sqrt{ 4 n_1^2 n_5^2 + m_1^2 r_e^4 } \right) \, ,
\end{align}
As a check, when $r_e = 0$, this reduces to the result we found for the extremal case:
\begin{align}
    \tilde{g} = \frac{n_1}{m_1} + \frac{n_5 \VT}{ m_1 } \, .
\end{align}

\subsubsection*{\ul{\it What happens for negative mass?}}

In our preceding discussion, we explicitly restricted to the case of $r_e^2 \geq 0$ which corresponds to a black hole solution. In $\AdS_3$, the spacetime mass can be negative. This includes global $\AdS_3$, which corresponds to the vacuum state of the holographic CFT, along with conical deficit spacetimes with energies below the $M=0$ BTZ solution. Let us explore what happens in this setting. To avoid confusion, let us parametrize the non-extremality factor as follows
\begin{align}
    f_e = 1- \frac{r_e^2}{u^2}= 1 + \frac{|r_e^2|}{u^2}\,, \qquad r_e^2<0\,.
\end{align}
Here $r_e^2$ parametrizes a negative mass and there is no horizon. Staring at the form of our solutions suggests an immediate issue. The combination $f_1f_5-f_e$ appears in the dilaton \C{nonextremaldilaton}, the metric \C{nonextremaldecoupled} and fluxes \C{finalquantfluxes}. However this combination now vanishes at some $u=r_c$ with value
\begin{align}
    r_c = \frac{r_1 r_5}{\sqrt{ |r_e^2| - r_1^2 - r_5^2 } } \, .
\end{align}
Fortunately we are saved from a potential singularity by the constraints \C{constraints}, 
\begin{align}
    | r_e^2 | -  r_1^2 - r_5^2 = - \frac{1}{2} \left( \sqrt{ 4 \tilde{g}^2 n_5^2 + r_e^4 } +  \sqrt{ \frac{4 \tilde{g}^2 n_1^2}{ \VT^2}  + r_e^4 }  \right) <0\,,
\end{align}
so $r_c$ is imaginary and the negative mass solutions appear to be well-behaved. 

Finally let us expand the metric \C{nonextremaldecoupled} converted to Einstein-frame near $u=0$, which gives 
\begin{align} \label{nonextremalads}
    \frac{\sqrt{\tilde{g}} }{\alpha'} ds^2&=  \sqrt{r_1 r_5^3} \left[  \frac{c_1 }{(r_1 r_5)^2} \left(- | r_e^2 |  d \tilde{t}^2 + u^2 d \tilde{x}_5^2 \right) 
     + \frac{1}{| r_e^2 | } \, du^2 +  \, d \Omega_3^2 \right]    + \sqrt{ \frac{r_1 \VT}{r_5} } \, \widetilde{ds}^2_{T^4} \, .
\end{align}
We again define a coordinate ${\hat \varphi} = \frac{\tilde{x}_5}{\tilde{R}}$ with periodicity $2\pi$. We want no conical excess for the metric, 
\begin{align}
    \frac{\sqrt{\tilde{g}} }{\alpha'} ds^2&=  \sqrt{r_1 r_5^3} \left[ \ldots + d\hat{u}^2 + \frac{c_1 \tilde{R}^2 | r_e^2 | }{(r_1r_5)^2} {\hat u}^2 d{\hat \varphi}^2  \right] + \ldots\,, 
\end{align}
which restricts the maximal value of $r_e$ to
\begin{align}
    | r_e^2 | \leq \frac{(r_1r_5)^2}{c_1 \tilde{R}^2} \,.
\end{align}
Note that $r_1$ and $r_5$ themselves are functions of $r_e$ from \C{constraints} and we have yet to determine $c_1$. At this stage, this is a more complicated relation than the usual relation leading to the mass of global $\AdS_3$. We will revisit this relation momentarily.

\subsubsection*{\ul{\it Fixing $c_1$ and a surprise involving $\tilde{g}$!}}

For cases with both positive and negative mass, we need to fix the constant $c_1$ in \C{nonextremaldecoupled}. First we restore the original definition of $f_e= 1 - \frac{r_e^2}{u^2}$ with positive $r_e^2$ corresponding to positive mass and negative $r_e^2$ corresponding to negative mass. 

We will use the following principle: at spatial infinity where $u\rightarrow \infty$, we would like the leading asymptotics of our solutions to be independent of the mass parameter $r_e$. Expanding the Einstein-frame metric near $u=\infty$ gives, 
\begin{align}
    \frac{1 }{\alpha'} ds_E^2 = \frac{\sqrt[4]{r_1^2 + r_5^2 + r_e^2}}{\sqrt{ \tilde{g} u}} \,du^2 + \ldots  = d\rho^2+\ldots\, ,
\end{align}
where $\rho = \frac{4}{3} u^{3/4} \, \tilde{g}^{-1/4} \left( r_1^2 + r_5^2 + r_e^2 \right)^{1/8}$. Writing the metric in terms of $\rho$ gives,
\begin{align}
    \frac{1 }{\alpha'} ds_E^2 = d\rho^2 +\frac{9c_1 \rho^2}{ 16( r_1^2 + r_5^2 + r_e^2 )} \left( - d \tilde{t}^2 + d \tilde{x}_5^2 \right) + \ldots \,.
\end{align}
We then choose to define
\begin{align}
    c_1 = ( r_1^2 + r_5^2 + r_e^2 ) \,,
\end{align}
which makes these leading metric terms independent of mass.

We have yet to examine the sphere metric, the torus metric and the dilaton at large $u$. In each case, the asymptotics involve the combination
\begin{align}
    \frac{\tilde g^2}{( r_1^2 + r_5^2 + r_e^2 )} \,,
\end{align}
which is not obviously mass independent based on our discussion so far. It is also not immediately mass independent when one examines the expressions for $r_1$ and $r_5$ in \C{constraints} and for $\tilde g$ in \C{solutiong}. However expanding the ratio in Mathematica in a power series and simplifying the coefficients gives mass independence to at least $(r_e)^{40}$ power. This is compelling evidence for the following relation, 
\begin{align} \label{magicrelation}
    \frac{\tilde{g}^2}{r_1^2 + r_5^2 + r_e^2} = \frac{\VT}{m_1}  \quad \Rightarrow \quad c_1 = \frac{m_1 }{\VT}\tilde{g}^2 \,,
\end{align}
which is a beautiful surprise! It means all the leading asymptotics are actually mass independent as we might have hoped. 
Let us spell this out. The Einstein-frame metric at large $\rho$ now takes the form, 
\begin{align} \label{nonextremalasymmetric}
  \frac{1}{\alpha'} ds^2_E =  d\rho^2 +\frac{9}{16} \rho^2 \left( - d \tilde{t}^2 + d \tilde{x}_5^2 \right) + \frac{9}{16}  \rho^2\, d \Omega_3^2 + \left( \frac{16}{9} \frac{m_1 \VT^{1/2}}{\rho^2} \right)^{1/3} \, \widetilde{ds}_{T^4} + \ldots\,,
\end{align}
with $\rho = \frac{4}{3} u^{3/4} \sqrt[8]{\frac{m_1}{\VT}}$. The dilaton takes the form, 
\begin{align}\label{nonextremalasymdilaton}
    \Phi = \frac{4}{3} \log ( \rho ) - \frac{2}{3} \log \left( \frac{16 m_1}{9 \VT}  \right) + O\left(\frac{1}{\rho^{8/3}}\right)\, .
\end{align}
The coefficient of the subleading term decaying with $\rho$ is mass-dependent. It is reasonable to use these asymptotics to define this theory of quantum gravity. 

\subsubsection*{\ul{\it Revisiting the most negative mass solution}}

Now we can wrap up a loose end involving the solution with largest negative mass. Specifically: does this solution become global $\AdS_3$ for sufficiently small $u$? We can check this in string-frame. Let us again define hatted variables as in \C{hatted}. Dropping the ``1''s in both $f_1$ and $f_5$ gives the metric:
\begin{align}
    {1\over \alpha'}ds^2 = \frac{c_1 {\tilde R}^2}{r_1 r_5} u^2 \left( - f_e d{\hat t}^2+d\hat{\varphi}^2\right) + \frac{r_1r_5}{u^2} \frac{1}{f_e} du^2 +\ldots.
\end{align}
We then define
\begin{align}
    u^2 = \frac{(r_1r_5)^2}{c_1 {\tilde R}^2}\hat{u}^2  
\end{align}
so the metric now takes the form
\begin{align}
    ds^2 = r_1r_5 \alpha'  \left[ \hat{u}^2 \left( - f_e d{\hat t}^2+d\hat{\varphi}^2\right) + \frac{1}{\hat{u}^2} \frac{1}{f_e} d\hat{u}^2 +\ldots \right]\,,
\end{align}
with $f_e = 1 + \frac{1}{{\hat u}^2}$ which is indeed the metric for global $\AdS_3$.

\subsection{Computing the mass} \label{sec:mass}

Ideally we would want to compute the mass of these spacetimes using the covariant phase space formalism. That is an interesting computation to which we hope to return elsewhere. Instead, we will use the structure of the undecoupled solutions that are asymptotically flat to ascribe a mass to each spacetime using the ADM formula. The positive mass theorem guarantees this ADM mass is positive, but it is reasonable to shift the value of the mass so that the extremal solution ($r_e=0$) has zero mass.  We will then study the decoupling limit applied to the ADM mass. 

\subsubsection*{\ul{\it The ADM mass}}

The solution generated by the procedure of section \ref{solutionsfromTST} prior to decoupling has a large $r$ $D=10$ asymptotic string-frame  metric read from (\ref{tststringframe}),
\begin{align} \label{leadingmetricundecoupled}
     \, ds^2 = -dt^2 + dx_5^2 + dr^2 + r^2 d\Omega_3^2 + ds^2_{T^4} + O\left(\frac{1}{r^2}\right)\,.
\end{align}
 To compute the ADM mass, we first reduce the string-frame metric from $D=10$ to $D=6$ on $T^4$. This introduces additional radial functions from the $T^4$ metric, which now multiply  $\sqrt{-g} e^{-2\Phi} R$. We then transform to Einstein-frame in $D=6$ and expand $g_{tt}$ to extract the subleading term, which takes the form: 
\begin{align} \label{einstein-string}
     g_{tt} &=  - 1 + \frac{\mu}{r^2} + \ldots \,, \\
    &= - 1 + \alpha'\, \frac{r_1^2 + r_5^2 + 3 r_e^2 + \left( r_1^2 + r_5^2 + r_e^2 \right) \cosh ( 2 \alpha ) }{4 r^2} + \ldots  \, ,
\end{align}
where $\mu$ is the mass parameter. What we are really describing is a black string in $D=6$ with a tension governed by $\mu$. 
Now we define $G_6$ via
\begin{align}
    \frac{1}{16\pi G_6} = \frac{1}{2 \kappa_6^2} = \frac{16\pi^4 \VT (\alpha')^2 }{( 2 \pi )^7 \alpha^{\prime 4} g^2}  \quad \Rightarrow \quad G_6 =  \frac{g^2 \pi^2 \alpha^{\prime 2} }{2 \VT} \, .
\end{align}
We note that 
\begin{align}
    \cosh(2\alpha) = \frac{2b^2}{(\alpha')^2} - 1\,, \quad r_5^2 = \frac{g\alpha'}{b} n_5\,, \quad r_1^2 = \frac{g \alpha'}{b \VT} n_1 \,.
\end{align}
Note that $\alpha\rightarrow 0$ corresponds to $b\rightarrow \alpha'$. In that limit, we should recover the usual formulae for the non-extremal D1-D5 system with no fundamental strings as a sanity check. We find the ADM black string tension,
\begin{align}
    T_{\rm ADM}(\alpha\rightarrow 0) \sim \frac{\mu}{G_6} = \frac{g ( n_1 + n_5 \VT ) + 2 r_e^2 \VT}{\pi^2 \alpha' g^2} \, ,
\end{align}
where we have ignored constants independent of mass.\footnote{The usual formula for the ADM mass in $D$ spacetime dimensions is $M_{\rm ADM} = \frac{(D-2)}{16 \pi G_D} \mathrm{vol} ( S^{D-2} ) \, \mu$.} For the extremal case ($r_e=0$) this gives, 
\begin{align}
    T_{\rm ADM}(\alpha\rightarrow 0) \sim \frac{\mu}{G_6} = \frac{ ( n_1 + n_5 \VT )}{ \pi^2  \alpha' g} \, ,
\end{align}
which agrees with our expectation of the tension of wrapped D1-branes and D5-branes. 

Finally we can go beyond our sanity check and compute the tension without taking $\alpha \rightarrow 0$. In this case we find
\begin{align}\label{MADM}
    T_{\rm ADM} \sim \frac{\mu}{G_6} = \frac{b^2 r_e^2 \VT + b g ( n_1 + n_5 \VT ) \alpha' + r_e^2 \VT \alpha^{\prime 2}}{\pi^2 g^2 \alpha^{\prime 3}} \, .
\end{align}
This ADM tension is expressed in terms of $b$, but $b$ is not an independent parameter of the problem. Rather, to determine $b$ in terms of the physical parameters of the problem we need to solve for string charge quantization again using the undecoupled solutions. This means imposing
\begin{align}
  \frac{1}{(2\pi)^6 (\alpha')^{3}}  \int_{S^3 \times T^4} \left( H_7 +\omega_7\right) = -m_1\, ,
\end{align}
which relates $b$ to $(m_1, n_1, n_5, g, \VT)$. This relation is quite involved for the non-extremal cases so we first turn to the extremal limit.

\subsubsection*{\ul{\it Extremal mass}}

Restricting to the extremal case first with $r_e=0$, we find the following expression for $b=b_{\rm ext}$, 
\begin{align} \label{bextremal}
    b_{\rm ext} = \frac{ \alpha' \sqrt{ g^2 m_1^2 + ( n_1 + n_5 \VT )^2} }{n_1 + n_5 \VT} \, ,
\end{align}
which gives the extremal tension
\begin{align}
    T_{\text{ADM}}^{\rm ext} \sim \frac{ \sqrt{ g^2 m_1^2 + ( n_1 + n_5 \VT )^2 } }{  \pi^2  \alpha' g} \, .
\end{align}
This is in perfect agreement with our expectation that the brane system is a bound state of $m_1$ fundamental strings, $n_1$ D1-strings and $n_5$ wrapped D5-branes. 

\subsubsection*{\ul{\it The almost extremal case}}

To solve for $b$ in the general case involves solving the following quartic equation: 
\begin{align} \label{exactb}
     \frac{ ( b^2 - \alpha^{\prime 2} ) \left( b  \VT r_e^2 + g \alpha' ( n_1 + n_5 \VT ) \right)^2 }{g^4 \alpha^{\prime 4}} =m_1^2 \, .
\end{align}
For small $r_e^2$ of either sign we can compute the leading correction to $b$ which takes the form, 
\begin{align}
    b =b_{\rm ext} \left( 1 - \frac{g m_1^2  \VT}{( n_1 + n_5 \VT)^2 \sqrt{ g^2 m_1^2 + ( n_1 + n_5 \VT)^2 }} r_e^2 \right) + \ldots \, .
\end{align}
Computing the change in the ADM tension $\Delta T_{\rm ADM} = T_{\rm ADM} -  T_{\text{ADM}}^{\rm ext}$ gives
\begin{align}
    \Delta T_{\rm ADM} = \frac{2 \VT}{ \pi^2 \alpha' g^2} r_e^2 +
    \ldots\, .
\end{align}

\subsubsection*{\ul{\it The extremely positive mass case}}

Now we turn to the case where $r_e^2$ is very large and positive. The equation \C{exactb} collapses to a quadratic at leading order
\begin{align}
    \frac{ ( b^2 - \alpha^{\prime 2} ) b^2  }{g^4 \alpha^{\prime 4}} (\VT r_e^2)^2 =m_1^2 \, ,
\end{align}
with
\begin{align}
    b^2 = \alpha^{\prime 2} \left(1 + \frac{m_1^2 g^4}{(\VT r_e^2)^2} \right) + O\left( \frac{1}{r_e^8}\right)\,.
\end{align}
This gives an ADM tension, 
\begin{align}
    T_{\rm ADM} \sim \frac{\mu}{G_6} = \frac{ 2 {\VT}}{ \pi^2 \alpha^{ \prime  } g^2 } r_e^2 +\ldots\, ,
\end{align}
where the omitted terms are suppressed by $r_e^2$. This is the typical dependence of the ADM energy on $r_e^2$.

\subsubsection*{\ul{\it Decoupling limit $r_e^2>0$}}

We want to study what happens to the energy in the decoupling limit. In this limit, we can simplify \C{exactb} by substituting $ g = \frac{{\tilde g} b}{\alpha'}$ and the equation becomes tractable with
\begin{align}
    b = \frac{\alpha' \left( r_e^2 \VT + \tilde{g} ( n_1 + n_5 \VT )\right)}{\sqrt{ r_e^4 \VT^2 + 2 \tilde{g} r_e^2 \VT ( n_1 + n_5 \VT ) + \tilde{g}^2 ( n_1 + n_5 \VT )^2 - \tilde{g}^4 m_1}} \, .
\end{align}
We now need to substitute for $\tilde{g}$ using \C{solutiong} then
when $r_e$ is very large, we see that
\begin{align} \label{largebdecoupling}
    b = \alpha' \left(  \sqrt{\frac{m_1}{m_1-1}} - \frac{ (n_1 + n_5 \VT )}{ ( m_1 - 1 )^{\frac{3}{2}}\sqrt{\VT}} \cdot\frac{1}{r_e} \right) + O \left( \frac{1}{r_e^2} \right) \, . 
\end{align}
Now we need to revisit the ADM energy. %
We first convert the ADM tension to a mass by wrapping the black string on the $x_5$ circle,  
\begin{align}
    M_{\rm ADM} =  T_{\rm ADM} (2\pi R) \,. 
\end{align}
This is still measured with respect to $G_6$ but that is not a good definition of the gravitational interaction strength in the decoupled region. Instead we want to measure the energy with respect to 
\begin{align} \label{decoupledG6}
    {\widetilde G}_6 = \frac{ \pi^2 \alpha^{\prime 2} }{2 \VT} \, ,
\end{align}
where we have stripped off the $g^2$ factor which will now depend on $r_e$ after decoupling. If we are fortunate, this should happen automatically when we evaluate the ADM mass expression in the decoupling limit. 

To take that limit, we first note that $t \rightarrow \sqrt{c_1} \frac{b}{\ell_s}\tilde{t} $ which means the mass becomes ${\tilde M} =  \sqrt{c_1}\frac{b}{\ell_s} M_{\rm ADM}$. Similarly $R \rightarrow \sqrt{c_1}\frac{b}{\ell_s} {\tilde R}$. Putting this together gives, 
\begin{align}
    M_{\rm ADM} \rightarrow {\tilde M} & = c_1 \frac{b^2}{\alpha'} (2\pi {\tilde R} ) \frac{\mu}{G_6} = \frac{4 m_1   }{\pi \alpha' } {\tilde R} \mu \,, \cr 
    &= 2 m_1 \tilde{R} \left( \frac{ r_e^2 \VT \alpha^{\prime 2} + b^2 \left( r_e^2 \VT +  \tilde{g} ( n_1 + n_5 \VT ) \right) }{\pi \VT \alpha^{\prime 2}} \right)\, .
\end{align}
So far we have not made any approximations but now we can substitute the leading large $r_e$ expression for $b$ from \C{largebdecoupling} and drop subleading terms in $r_e$ to find:
\begin{align} \label{decoupledmass}
    {\tilde M} =   \frac{2m_1(2 m_1 - 1)}{\pi (m_1 - 1)} \tilde{R} \, r_e^2 + O\left(r_e\right)\, .
\end{align}

\subsection{A look at the entropy}

Lastly, we can examine the entropy for these large black holes by computing the horizon area.

\subsubsection*{\ul{\it Undecoupled case}}

We use the Einstein-frame metric obtained from \C{nonextremalundecoupled} using \C{undecouplednonextremaldilaton} to compute, 
\begin{align}
    S = \frac{A}{4 G_6} & = \frac{(2\pi R) {\rm vol}(S^3) \ell_s^3 r_e^3 \cosh ( \alpha ) }{4 G_6} + \ldots \,,  \nonumber \\
    &= \frac{2 \pi R \VT \cosh ( \alpha ) }{g^2 \ell_s} r_e^3 + \ldots \, ,
\end{align}
where we have taken $r_e$ large. This is the expected result for black holes in $D=5$ asymptotically flat spacetime. 

\subsubsection*{\ul{\it Decoupled case}}

Now we turn to the decoupled case with string-frame metric \C{nonextremaldecoupled} and dilaton \C{nonextremaldilaton}. We use the gravitational coupling \C{decoupledG6} which gives the entropy, 
\begin{align}
    S = \frac{A}{4 {\widetilde G}_6} &= \frac{ \frac{( 2 \pi \tilde{R} ) m_1 \mathrm{vol} ( S^3 ) \alpha^{\prime 2} r_e^2}{\VT} }{ 4 \cdot \frac{\pi^2 \alpha^{\prime 2}}{2 \VT} } = 2 \pi \tilde{R} m_1 r_e^2 + \ldots\, ,
\end{align}
where we have again assumed $r_e$ large and omitted suppressed terms. If we write this in terms of the mass ${\tilde M}$ of \C{decoupledmass}, we see that
\begin{align} \label{hagedorn}
    S({\tilde M}) = \frac{ ( m_1 - 1 ) }{2 m_1 - 1} \pi^2 \tilde{M} +\ldots \, .
\end{align}
This implies a Hagedorn density of states with an associated Hagedorn temperature; see~\cite{Fichet:2023xbu} for a bottom up analysis of black holes in linear dilaton spacetimes. The precise value of the Hagedorn temperature depends on the numerical factors in the definition of the ADM mass but that the growth rate is Hagedorn appears to be robust.

\section*{Acknowledgements}

We would like to thank Iosif Bena and Troels Harmark for helpful correspondence, and Emil Martinec for very helpful comments on the manuscript. We also thank the organizers of the MATRIX Program in Melbourne, AU ``New Deformations of Quantum Field and Gravity Theories’’ (January 2024) and the Simons Workshop on ``Exact approaches to low-supersymmetry AdS/CFT'' (October 2024). C.~F. is grateful to the organizers of the ``Workshop on Higher-$d$ Integrability'' in Favignana, Italy (June 2025), where part of this work was performed. 
S.~S. similarly thanks the organizers of the Simons Workshop on ``Landscapia'' (August 2024) and ``Lotus and Swamplandia'' (June 2025).

C.~F. is supported by U.S. Department of Energy grant DE-SC0009999, funds from the University of California, and the National Science Foundation under Cooperative Agreement PHY-2019786 (the NSF AI Institute for Artificial
Intelligence and Fundamental Interactions).
S.~S. is supported in part by NSF Grant No. PHY2014195 and NSF Grant No. PHY2412985.

\newpage
\appendix

\section{Flux Quantization}\label{app:flux}

Let us summarize the situation with flux quantization in type IIB string theory with no additional sources like D-branes. The quantization condition depends on the definition of the flux and its Bianchi identity. 
Consider the type IIB supergravity action in string-frame, 
\begin{align}\label{SUGRAIIB}
    S_{{\rm IIB}} = {1\over 2 \kappa_{10}^2}\int d^{10}x \, \sqrt{- g} \Bigg\{ & e^{-2\Phi} \left( R + 4 (\p \Phi)^2 - {1\over 12} |H_3|^2\right) - {1\over 2} |F_1|^2 \cr & - {1\over 12}|F_3 - C_0 H_3|^2+ \ldots \Bigg\}.
\end{align}
The field $C_0$ is periodic with $C_0 \sim C_0+1$, and the complex type IIB string coupling is $\tau = C_0 + i e^{-\Phi}$. 

\subsubsection*{\ul{\it Quantization of RR charges}}

Now let's turn to the RR sector. There are subtleties in quantizing RR fields because of the self-duality condition and the Bianchi identity \C{RRbianchi}. However the non-closure of the RR flux is always proportional to $H_3$. In our solutions $H_3$ is only non-vanishing in the spacetime directions. 

For this reason we can assume the RR flux is a closed form in the directions that matter for quantization, 
and assume the simplest quantization condition for fluxes supported along $\S_p \subset S^3 \times T^4$: 
\begin{align} \label{Fquant}
 \frac{1}{(2\pi \sqrt{\a'})^{p-1}}   \int_{\S_p} { F}_p  \in \Z \, , \qquad { F}_p = dC_{p-1} \, .
\end{align}
This applies to $F_3$ and $F_7$. 

\subsubsection*{\ul{\it Quantization of NS5-brane charge}}

The quantization condition from the string world-sheet is insensitive to RR fluxes and follows from the type II string Bianchi identity $dH_3=0$,
\begin{align}
{1\over 4\pi^2 \alpha'} \int_{\S^3} H_3 = m_5\, ,   
\end{align}
where $m_5$ is an integer for any compact $\S_3$. The RR field strengths satisfy, 
\begin{align} \label{RRbianchi}
    dF_p  + H_3 \wedge F_{p-2} =0\, , 
\end{align}
and the $B_2$-field equation of motion gives, 
\begin{align}
    d(e^{-2\phi} \ast H_3 ) - F_1 \wedge F_7 + F_3 \wedge F_5 = 0 \, , \label{Beom}
\end{align}
with $\ast F_5 = F_5$ and $\ast F_3 = - F_7$. We want to rewrite this in the form $dH_7=0$. This is possible because \C{Beom} is closed using the Bianchi identities. Let us use the trivialization,
\begin{align}
    F_p = dC_{p-1} + H_3 \wedge C_{p-3}. 
\end{align}

\subsubsection*{\ul{\it Quantization of string charge}}

The final issue we will address is the quantization of $H_7 = \ast e^{-2\phi} H_3$ which determines fundamental string charge. From \C{Beom}, we see that 
\begin{align}
    d H_7 = F_1 \wedge F_7 - F_3 \wedge F_5 \, . \label{bianchh7}
\end{align}
We would like to trivialize the right side of \C{bianchh7},
\begin{align}
    d \omega_7 = F_3 \wedge F_5 - F_1 \wedge F_7 \, ,
\end{align}
for some $\omega_7$, so that $d ( H_7 + \omega_7 ) = 0$. The relations we will need are,
\begin{align}
    d F_p = - H_3 \wedge F_{p-2}\,,
\end{align}
along with the choice of trivializaton
\begin{align}
    F_p = d C_{p-1} + H_3 \wedge C_{p-3} \, .
\end{align}
We proceed by enumerating all terms that can appear, organized by the number of $B_2$ factors, and fixing coefficients.

The terms containing one Ramond flux and one Ramond potential but with no $B_2$ are
\begin{align}
    a_1 C_0 F_7 + a_2 C_2 \wedge F_5 + a_3 C_4 \wedge F_3 + a_4 C_6 \wedge F_1  \, ,
\end{align}
where the $a_i$ are undetermined coefficients. Acting with $d$ on each term, we find
\begin{align}
    d \left( a_1 C_0 F_7 \right) &= a_1 F_1 \wedge F_7 + a_1 C_0 d F_7 \nonumber \\
    &= a_1 F_1 \wedge F_7 - a_1 C_0 H_3 \wedge F_5 \, , \nonumber \\
    d \left( a_2 C_2 \wedge F_5 \right) &= a_2 d C_2 \wedge F_5 + a_2 C_2 \wedge d F_5 \nonumber \\
    &= a_2 \left( F_3 - C_0 H_3 \right) \wedge F_5 - a_2 C_2 \wedge H_3 \wedge F_3 \nonumber \\
    &= a_2 F_3 \wedge F_5 - a_2 C_0 H_3 \wedge F_5 - a_2 C_2 \wedge H_3 \wedge F_3 \, , \nonumber \\
    d \left( a_3 C_4 \wedge F_3 \right) &= a_3 d C_4 \wedge F_3 + a_3 C_4 \wedge d F_3 \nonumber \\
    &= a_3 \left( F_5 - H_3 \wedge C_2 \right) \wedge F_3 - a_3 C_4 \wedge H_3 \wedge F_1 \nonumber \\
    &= a_3 F_5 \wedge F_3 - a_3 H_3 \wedge C_2 \wedge F_3 - a_3 C_4 \wedge H_3 \wedge F_1 \, , \nonumber \\
    d \left( a_4 C_6 \wedge F_1 \right) &= a_4 d C_6 \wedge F_1 \nonumber \\
    &= a_4 \left( F_7 - H_3 \wedge C_4 \right) \wedge F_1 \nonumber \\
    &= a_4 F_7 \wedge F_1 - a_4 H_3 \wedge C_4 \wedge F_1 \, .
\end{align}
We demand that the terms proportional to $F_1 \wedge F_7$ have overall coefficient $-1$, which gives the constraint
\begin{align}
    a_1 - a_4 = - 1 \, .
\end{align}
We also require that the terms proportional to $F_3 \wedge F_5$ have coefficient $+1$, which gives
\begin{align}
    a_2 - a_3 = 1 \, .
\end{align}
There are still two free coefficients after imposing these relations.

The remaining terms in $d \omega_7$ which we need to eliminate are
\begin{align} \label{mustkill}
   - \left( a_1 + a_2 \right) C_0 H_3 \wedge F_5 - \left( a_2 + a_3 \right) C_2 \wedge H_3 \wedge F_3 - \left( a_3 + a_4 \right) C_4 \wedge H_3 \wedge F_1 \, .
\end{align}
We can solve this by setting
\begin{align} \label{generalsolutionfora}
    ( a_1, a_2, a_3, a_4 ) = \frac{1}{2} \left( -1, 1, -1, 1 \right) \, .
\end{align}
However, this solution is not unique if some of the wedge products in \C{mustkill} vanish. In our specific model, the only non-vanishing $H_3$ is along the spacetime directions and so no terms in \C{mustkill} are along $S^3\times T^4$. We therefore have more freedom in choosing $( a_1, a_2, a_3, a_4 )$. Using this freedom we can choose $a_1=-1, a_2=0, a_3=-1, a_4=0$. Thus our ansatz for $\omega_7$ is
\begin{align} \label{defomega7}
    \omega_7 = -\left( C_0 F_7  + C_4 \wedge F_3 \right) \, .
\end{align}
In particular,
\begin{align}
    d \left( \omega_7 + H_7 \right) = 0 \, .
\end{align}

Since $H_7$ is gauge invariant, this relation guarantees that $d \omega_7$ is gauge-invariant. Therefore, under any gauge transformation, $\omega_7$ may shift by at worst a closed term. Let us verify this in some explicit cases for the general solution \C{generalsolutionfora}. The $C_6$ gauge transformation is simplest, since
\begin{align}
    C_6 \to C_6 + d \Lambda_5 \, , 
\end{align}
and none of the other Ramond potentials appearing in $\omega_7$ are affected. Under such a gauge transformation, we have
\begin{align}
    \delta \omega_7 = \frac{1}{2} d \Lambda_5 \wedge F_1 \, = \frac{1}{2} d \left( \Lambda_5 \wedge F_1 \right) \, .
\end{align}
So in this case $\delta \omega_7$ is exact.

Let us now consider gauge transformations for the lower Ramond potentials, which mix with the higher ones. If we transform $C_4$ as
\begin{align}
    C_4 \to C_4 + d \Lambda_3 \, , 
\end{align}
then since the combination
\begin{align}
    F_7 = d C_6 + H_3 \wedge C_4
\end{align}
is gauge invariant, we should have
\begin{align}
    \delta \left( d C_6 \right) = - H_3 \wedge \delta C_4 = - H_3 \wedge d \Lambda_3 \, .
\end{align}
Therefore we must also shift $C_6$ as
\begin{align}
    \delta C_6 = H_3 \wedge \Lambda_3 \, .
\end{align}
Now the combined transformation of $\omega_7$ is
\begin{align}
    \delta \omega_7 &= \frac{1}{2} \left( - \delta C_4 \wedge F_3 + \delta C_6 \wedge F_1 \right) \nonumber \\
    &= \frac{1}{2} \left( - d \Lambda_3 \wedge F_3 + H_3 \wedge \Lambda_3 \wedge F_1 \right) \nonumber \\
    &= - \frac{1}{2} \left( d \Lambda_3 \wedge d C_2 + C_0 d \Lambda_3 \wedge H_3  + \Lambda_3  \wedge H_3 \wedge F_1 \right) \nonumber \\
    &= - \frac{1}{2} d \left( \Lambda_3 \wedge d C_2 + C_0 \Lambda_3 \wedge H_3 \right) \, ,
\end{align}
and again $\omega_7$ shifts only by an exact quantity.

Similarly, suppose that we perform a $C_2$ gauge transformation
\begin{align}
    C_2 \to C_2 + d \Lambda_1 \, , 
\end{align}
under which we must likewise have $\delta F_5 = \delta \left( d C_4 + H_3 \wedge C_2 \right) = 0$, which means
\begin{align}
    \delta \left( d C_4 \right) = - H_3 \wedge \delta C_2 = - H_3 \wedge d \Lambda_1 \, , 
\end{align}
so we transform
\begin{align}
    C_4 \to C_4 + H_3 \wedge \Lambda_1 \, .
\end{align}
Now $C_6$ will not shift because $\delta C_4 \wedge H_3 = 0$. The combined variation of $\omega_7$ is
\begin{align}
    \delta \omega_7 &= \frac{1}{2} \left( d \Lambda_1 \wedge F_5 - H_3 \wedge \Lambda_1 \wedge F_3 \right) \nonumber \\
    &= \frac{1}{2} \left( d \Lambda_1 \wedge d C_4 + d \Lambda_1 \wedge H_3 \wedge C_2 + \Lambda_1 \wedge H_3  \wedge d C_2 \right) \nonumber \\
    &= \frac{1}{2} d \left( \Lambda_1 \wedge d C_4 + \Lambda_1 \wedge H_3 \wedge C_2 \right) \, ,
\end{align}
and once again $\delta \omega_7$ is exact.

Finally, under a shift $C_0 \to C_0 + \epsilon$, it is \emph{not} the case that $\omega_7$ shifts by an exact form. But this is acceptable, since only the integer shifts $C_0 \to C_0 + n$ for $n \in \mathbb{Z}$ should be genuine symmetries of the string theory background, and these are large gauge transformations (in the sense that they are not continuously connected to the identity). We expect that the Page charge \cite{Page:1983mke} (for a discussion, see \cite{Marolf:2000cb}) should not be invariant under such large gauge transformations, but only under small ones.

Finally we impose the quantization condition
\begin{align} \label{H7quant}
  \frac{1}{(2\pi)^6 (\alpha')^{3}}  \int_{\S_7} \left( H_7 +\omega_7\right) = -m_1\, ,
\end{align}
through any compact $\S_7$. This deals with the NS sector flux quantization in a general setting. We are interested in the specific case of a spacetime $\M_3 \times S^3 \times T^4$ so only components of $H_7 +\omega_7$ along $S^3\times T^4$ are quantized.

\section{T-duality Rules}\label{app:buscher}
In this Appendix we summarize the T-duality transformation properties of the metric, dilaton, $B_2$ potential and the various RR field strengths under duality in the $x$-direction \cite{Buscher:1987sk}: 
\begin{align}\begin{split}
	& e^{2 \Phi'} = \frac{e^{2 \Phi}}{G_{xx}} \, , \\
    & G_{xx}' = \frac{1}{G_{xx}} \, , \\
   & G_{\mu x}' = \frac{B_{\mu x}}{G_{xx}} \, , \\
    & G_{\mu \nu}' = G_{\mu \nu} - \frac{G_{\mu x} G_{\nu x} - B_{\mu x} B_{\nu x}}{G_{xx}}  \, , \\
   & B_{\mu x}' = \frac{G_{\mu x}}{G_{xx}} \, , \\
   & B_{\mu \nu}' = B_{\mu \nu} - \frac{B_{\mu x} G_{\nu x} - G_{\mu x} B_{\nu x}}{G_{xx}} \, , \\
   & F^{(p) \prime}_{\mu_1 \cdots \mu_{p-1} x} = F_{\mu_1 \cdots \mu_{p-1}}^{(p-1)} - (p - 1) \frac{ F_{[ \mu_1 \cdots \mu_{p-2} | x |}^{(p-1)} G_{\mu_{p-1} ] x }}{G_{xx}} \, ,\nonumber \\
 & F^{(p) \prime}_{\mu_1 \cdots \mu_p} = F_{\mu_1 \cdots \mu_p x}^{(p+1)} + p F_{[\mu_1 \cdots \mu_{p-1}}^{(p-1)} B_{\mu_p ] x} + p ( p - 1 ) \frac{F^{(p-1)}_{[ \mu_1 \cdots \mu_{p-2} | x |} B_{ \mu_{p-1} | x |} G_{\mu_p ] x } }{G_{xx}} \, .  
\end{split}\end{align}

\newpage
\bibliographystyle{utphys}
\bibliography{master}

\end{document}